\documentclass[runningheads]{llncs}
\pdfoutput=1
\usepackage{amsmath}
\usepackage{graphicx}
\usepackage{amssymb}
\usepackage{algorithm}
\usepackage{algpseudocode}

\usepackage{subfigure}
\usepackage{epstopdf}

\begin{document}
\title{FAST: A Fairness Assured Service Recommendation Strategy Considering Service Capacity Constraint}
\titlerunning{FAST: A Fairness Assured Service Recommendation Strategy}
%
\author{Yao Wu\inst{1} \and
Jian Cao\inst{1}\thanks{Corresponding author} \and
Guandong Xu\inst{2}}

\authorrunning{Y. Wu et al.}
%
\institute{Shanghai Jiao Tong University, Shanghai, China
\email{\{wuyaoericyy,cao-jian\}@sjtu.edu.cn}\\ \and
University of Technology Sydney, Sydney, New South Wales, Australia\\
\email{Guandong.Xu@uts.edu.au}}
\maketitle              
\begin{abstract}
An excessive number of customers often leads to a degradation in service quality. However, the capacity constraints of services are ignored by recommender systems, which may lead to unsatisfactory recommendation. This problem can be solved by limiting the number of users who receive the recommendation for a service, but this may be viewed as unfair. In this paper, we propose a novel metric \textit{Top-N Fairness} to measure the \textit{individual fairness} of multi-round recommendations of services with capacity constraints. By considering the fact that users are often only affected by top-ranked items in a recommendation, \textit{Top-N Fairness} only considers a sub-list consisting of top \textit{N} services. Based on the metric, we design \textit{FAST}, a \textit{F}airness \textit{A}ssured service recommendation \textit{ST}rategy. \textit{FAST} adjusts the original recommendation list to provide users with recommendation results that guarantee the long-term fairness of multi-round recommendations. We prove the convergence property of the variance of \textit{Top-N Fairness} of $FAST$ theoretically. \textit{FAST} is tested on the Yelp dataset and synthetic datasets. The experimental results show that \textit{FAST} achieves better recommendation fairness while still maintaining high recommendation quality. 
\end{abstract}
\section{Introduction}
In service recommendation, a user's degree of satisfaction with an item is affected by many factors. \textit{Capacity constraints}, which affect many service recommendation scenarios like dining, accommodation, fitness, haircuts, massages, medical services and so on, is a special factor that decides how many customers can receive a service with an assured level of quality. For example, a restaurant often has a capacity constraint on the number of customers who can be served during their dining hours. If too many customers arrive at a restaurant, their dining experience will be unpleasant or in the worst case, some customers will be very disappointed. However, recommender systems make recommendations to customers which only align with their preferences, and this may lead to dissatisfaction.

The solution of previous studies about recommendation with capacity constraint was to recommend the service to a limited number of users, or to penalize the service's relevant score when the recommended users exceeded the service's capacity \cite{christakopoulou2017recommendation}, making it less likely to be recommended. But such an approach brings a new problem, namely it is unfair to those users who may also like this restaurant according to their preference information. 

Fairness is already a concern in recommendation algorithm design \cite{stratigi2017fairness}. At present, the research on recommendation fairness mainly considers \textit{group fairness} \cite{zehlike2017fa}, trying to eliminate the influence of specific group attributes on the recommendation results, or removing the difference in recommendation results between groups caused by data bias, such as ensuring that gender or nationality does not affect the recommended results. Unlike these studies, the fairness we consider here is \textit{individual fairness} \cite{zafar2017fairness}, which ensures the same quality of recommendation for different users when capacity constraints are considered. To the best of our knowledge, we are the first to formalize the fairness ensured recommendation problem for services with capacity constraints. The contributions of this paper are as follows:
\begin{itemize}
	\item We propose a metric \textit{Top-N Fairness} to measure the fairness of recommendation under capacity constraint.
	\item We design a strategy named \textit{FAST} (\textit{F}airness \textit{A}ssured service recommendation \textit{ST}rategy) to ensure the long-term fairness of multi-round recommendations. We also prove the convergence property of the variance of \textit{Top-N Fairness} of \textit{FAST} theoretically. 
	\item Experiment results on a real-world dataset and synthetic datasets show \textit{FAST} can achieve higher fairness compared with baseline methods while still preserving high recommendation quality.
\end{itemize}
The rest of the paper is organized as follows. Section 2 discusses the related work. Section 3 formalizes the fairness assured multi-round recommendation problem for services with capacity constraints. Section 4 presents the fairness assured service recommendation strategy. The experiment results are illustrated in Section 5. We conclude the paper in Section 6.

\section{Related Work}
Many researchers have begun to focus on metrics other than recommendation accuracy to measure the performance of recommender systems \cite{ge2010beyond,pu2011user}, and fairness is one of the important metrics.

Currently, the research on fairness in recommender system can be roughly divided into two categories: \textit{group fairness} and \textit{individual fairness}. Ensuring \textit{group fairness} requires that the attributes of a specific group will not affect the recommendation results so the disadvantaged group can be given the same opportunities as the superior group \cite{zehlike2017fa,asudeh2019designing,beutel2019fairness,yao2017beyond,zhu2018fmsr}. Geyik et al. \cite{geyik2019fairness} proposed a re-ranking algorithm that reorders the results based on the recommended scores so that the distribution of the results meets the proportion of specific parameters; and Bose et al. \cite{bose2019compositional} tried to remove information about protected sensitive attributes in graph embedding by learning a series of adversarial filters. These approaches designed to ensure \textit{group fairness}  usually can only guarantee fair treatment in terms of one or a few attributes. However, when fairness in terms of some attributes is guaranteed, the unfairness related to other attributes may not be avoidable. 

Our approach focuses on the \textit{individual fairness} level, which has been considered by very few researchers. This metric puts emphasis on the view that similar users should be treated similarly \cite{geyik2019fairness,biega2018equity}. Rastegarpanah et al. \cite{rastegarpanah2019fighting} improved fairness by generating antidote data. The \textit{individual fairness} to be maintained by the recommender system in their work was inspired by \cite{zafar2017fairness}, and is defined as the equality of users' prediction accuracies. Our definition of \textit{individual fairness} is similar to theirs, but our interpretation of equal quality is different, which is the extent to which the final recommendation results considering fairness match the initial recommendation results considering a user's preference should be equal between users. 

Some research classifies approaches considering fairness from other perspectives. \cite{burke2017multisided} divided fairness-related criteria into consumers (C-fairness), providers (P-fairness) \cite{mehrotra2018towards,qian2015scram}, and both (CP-fairness) \cite{patro2020incremental} according to the stakeholders that systems consider; \cite{zafar2017fairness} classified approaches from the perspective of the time that the mechanism works in the system, and divided the fairness mechanism into pre-processing \cite{calmon2017optimized,kamiran2012data}, in-processing \cite{agarwal2018reductions,bose2019compositional} and post-processing \cite{liu2018personalizing,karako2018using} approaches. Our study considers consumer fairness and proposes a post-processing approach that further processes the existing recommendation results to obtain results that ensure individual fairness.

\section{Fairness Assured Multi-round Recommendation Problem for Services with Capacity Constraints}
We suppose that there is a conventional recommendation algorithm in the system, which provides a predicted rating matrix $R$ and the original recommendation lists $L$ for all users based on $R$. If we push $L$ directly to users, there is a high possibility that it will break the capacity constraints $C$ of services. To solve this problem, we design a strategy to adjust $L$ and generate new recommendation lists $L^T$ which can make recommendations as fair as possible without breaking capacity constraints while still preserving recommendation quality.

\subsection{Notations}
We use the following notations:
\begin{itemize}
	\item $S = \{s_1, s_2,..., s_m\}$ is a set of recommended services.
	\item $C = \{c_1, c_2,..., c_m\}$ is a set of services' capacity constraints.
	\item $U = \{u_1, u_2,..., u_n\}$ is a set of users.
	\item $R = \left[r_{1,1}, r_{1,2},..., r_{n,m}\right]$ is a relevant rating matrix produced by the original recommendation algorithm of the system.
	\item $L = \{l_1, l_2,..., l_n\}$ is a set of original recommendation lists based on $R$.
	\item $L^T = \{l_1^T, l_2^T,..., l_n^T\}$ is a set of recommendation lists finally outputted to users in the $T^{th}$ round recommendation.
	\item $\delta^T_i$ is a variable to indicate whether user $u_i$ uses the recommender system in the $T^{th}$ recommendation or not, where $\delta^T_i=1$ denotes yes and $\delta^T_i=0$ denotes no.
\end{itemize}

\subsection{Capacity Constraints}
We suppose each recommended service $s_j$ has a capacity constraint $c_j$. When there are multiple customer channels for a service, we only consider customers from the recommender system under discussion, as do the capacity constraints. In order to simplify the representation, unless otherwise specified, the capacity constraint of a service in our paper refers to the limited number of users to whom the recommendation system can recommend this service. This data can be obtained by dividing the allowed service capacity for the recommender system by the conversion rate of recommendation, which is the attendance ratio of users who are recommended a service.

\subsection{Recommendations on Top-$N$ Services}\label{Considering Top-N}
In practice, the recommendation results are shown in a limited space, such as web pages or APPs. Although longer lists can be shown to customers by pagination, research on user behaviors shows that most users only look at a few results before deciding, and they are more likely to notice highly ranked results \cite{joachims2007evaluating}. According to some reports, even an item at position 5 is largely ignored \cite{joachims2007search}. Therefore, we choose to ignore the influence of services in lower positions. To be more specific, we only consider the influence of a service's capacity constraint or a user's fairness when this service appears in the user's original top-$N$ service recommendation list. 

We denote the sub-list of the top \textit{N} services of user $u_i$'s original recommendation list as $l(N)_i$. At the same time, for each recommended service $s_j$, users whose $l(N)_i$ include $s_j$ are denoted as a user set $U_j$, and the number of users in  $U_j$ is often greater than the capacity constraint of $s_j$.

\subsection{Measuring the Fairness of Recommendations}\label{Different Fair Status}
Based on intuition, we divide the fairness status of a recommender system into three levels, and the fairness metric we design should be able to express these three levels of fairness:
\begin{enumerate}
	\item \textit{Perfect Fair Status}: At this level, every service in  a user's recommendation list is fairly recommended. Each user reaches an absolutely fair status on every service in his recommendation list and also an absolutely fair status of the whole top-$N$ service recommendation list.
	\item \textit{Individual Level Fair Status}: Each service is not necessarily allocated fairly, but each user can reach a fair status according to the top-$N$ service recommendation list. In this case, a user may lose the chance of being allocated some services in his recommendation list, but he has more chance than the others on other services, and the summation of the fairness degree on the top-$N$ service recommendation list offsets these deviations against each other, thereby achieving a fair status at an individual level. 
	\item \textit{Relatively Fair Status}: The system cannot ensure that every user reaches an absolutely fair status, but the degree of unfairness among the users is the same, thus achieving a relatively fair status at an individual level.
\end{enumerate}

Obviously, for a single round recommendation, it is unlikely to ensure fairness for all users due to capacity constraints. Instead, we measure the long-term fairness in the multi-round recommendation process in which users' fairness can accumulate over recommendations. We define two kinds of appearance probabilities of service to measure the chance of a service being allocated to users.

\begin{definition}[Overall Appearance Probability]
	The probability of a service $s_j$ appearing in the recommendation lists of all users in $U_j$ up to $T^{th}$ round recommendation is:
	\begin{equation}\label{E2}
	p^{T}_j = \frac{\sum_{u_i\in U_j}\sum^T_{t=0}\delta_i^t\cdot In\_tn(s_j,l_i^t,N)}{\sum_{u_i\in U_j}\sum^T_{t=0}\delta_i^t}
	\end{equation}
\end{definition}

\begin{equation}
In\_tn(s_j,list,N)=\left\{
\begin{aligned}
&0\text{ if $s_j$ is not in the top $N$ sub-list of $list$} \\
&1\text{ if $s_j$ is in the top $N$ sub-list of $list$}
\end{aligned}
\right.
\end{equation}

\begin{definition}[Actual Appearance Probability]
	The probability of a service $s_j$ appearing in the recommendation lists of user $u_i$ up to $T^{th}$ round recommendation is:
	\begin{equation}\label{E3}
	p_{i,j}^T = \frac{\sum^T_{t=0}\delta_i^t\cdot In\_tn(s_j,l_i^t,N)}{\sum^T_{t=0}\delta_i^t}
	\end{equation}
\end{definition}

If $u_i$ receives a fair recommendation on $s_j$, the \textit{Actual  Appearance Probability} of $s_j$ to $u_i$ should be equal to the \textit{Overall  Appearance Probability} of $s_j$. Thus, the difference between the above two appearance probability values represents the fairness degree of user $u_i$ on service $s_j$. Furthermore, we divide the difference by \textit{Overall Appearance Probability} to smooth the difference in capacity conflicts between different services and obtain the following definition:
\begin{definition}[Service Fairness Degree] Fairness degree of user $u_i$ on service $s_j$ up to $T^{th}$ round recommendation: 
	\begin{equation}\label{E4}
	F_{i,j}^T = \frac{p_{i,j}^T-p^{T}_j}{p^{T}_j}
	\end{equation}
\end{definition}

If $F_{i,j}^T$ is greater than zero, it means $u_i$ is allocated to service $s_j$ more frequently than the others in $U_j$; If $F_{i,j}^T$ is less than zero, service $s_j$ appears in his recommendation lists with fewer opportunities than the others; If $F_{i,j}^T$ is equal to zero, it means $u_i$ receives a fair recommendation for $s_j$. At the same time, we can add the service fairness degrees of all the services in user's $l(N)_i$ list and obtain the overall fairness degree at an individual level. We call it the fairness degree of Top-N recommendation(or \textit{Top-N Fairness} for short):
\begin{definition}[Top-N Fairness] Overall fairness degree of user $u_i$ up to $T^{th}$ round recommendation:
	\begin{equation}\label{E5}
	F_i^T = \sum_{s_j\in l(N)_i }F_{i,j}^T
	\end{equation}
\end{definition}

With the measurement of fairness, we can represent three levels of fairness in a formalized way:
\begin{enumerate}
	\item \textit{Perfect Fair Status}: $\forall u_i\in U$ and  $\forall s_j\in l(N)_i$, $F_{i,j}^T = 0$. 
	\item \textit{Individual Level Fair Status}: $\forall u_i\in U, F_i^T = 0$.
	\item \textit{Relatively Fair Status}: $\forall u_i,u_j\in U, F_i^T = F_j^T$. 
\end{enumerate}

These three fairness statuses share the same feature, the variance of \textit{Top-N fairness} among users is equal to zero. Therefore, we use the \textbf{variance among users' \textit{Top-N Fairness}} $D(F_i^T)$ as a measure of the fairness of recommender systems, where the smaller the variance, the fairer the recommender system.

\subsection{Quality of Recommendations}
When the original recommendation lists $L$ are adjusted to generate new recommendation lists $L^T$, those services whose capacity is constrained are removed from the list, hence the recommendation list cannot fully meet the users' preferences, and the quality of the recommendation list decreases. 

Following the idea in Section \ref{Considering Top-N}, we only consider the quality of $l(N)_i$. In the new list $l_i^T$, quality declines when a service is removed, and we use the predicted rating score of the service as a measure of the degree of decline. Therefore, the quality of a new recommendation list $l_i^T$ will be the sum of the rating scores of all the services belonging to $l(N)_i$ and $l_i^T$ at the same time. Taking into account that users may have different rating habits, i.e., some users like to give positive reviews while others prefer bad reviews, we use the highest rating of $l(N)_i$ as the denominator to normalize the quality score. Moreover, the positions of services in the recommendation lists also reflect their importance to a user. The quality measurement can be extended by giving each service a logarithmic discount based on its position in $l(N)_i$. The quality of the recommendation lists of the entire system can be obtained by adding the recommendation list quality of each user.

\begin{definition}[Quality of Recommendation List] Quality of outputted recommendation list $l_i^T$ of user $u_i$ on $T^{th}$ round recommendation:
	\begin{equation}
	q^T_i = \frac{\sum_{s_j\in l(N)_i \cap l(N)_i^t}\frac{r_{i,j}}{log_2(p_{i,j}^T+1)}}{r_{i,l(N)_i[0]}}
	\end{equation}
	where $l(N)_i[0]$ represents the subscript index of the service appearing at the top position of $l(N)_i$, and $p_{i,j}^T$ is the position of service $s_j$ in $l(N)_i$.
\end{definition}

Our problem is to generate $L^t$ based on a strategy so that the capacity constraint will not be violated while the fairness of the recommendation quality of each user can be assured along with an increase in recommendation times.

\section{A Fairness Assured Service Recommendation Strategy}
 When only fairness is considered, the problem can be reduced to a Knapsack problem which has been proven to be a non-deterministic polynomial complete problem. We analogize the capacity constraint of services as the capacity of the knapsack, the users' recommendation lists as the items put in the knapsack, and fairness as the objective. When further taking the quality of recommendation lists into consideration, the problem becomes more complicated. So we choose heuristic strategies to solve the problem.

In practical applications, service recommendations can be divided into two scenarios. The first is service recommendation for a fixed user set where the list of users receiving recommendations remains basically unchanged for a period of time, like an active advertising push, recommendations for members, high-end service recommendations, etc. This ensures the recommendation process in a stable environment, and the fairness of users is fully accumulated. The second is for a dynamic user set in which not all users receive recommendations each round or new users join, like recommendations for dining, movies to watch, medical services, etc. By considering the above situations, we design two versions of \textit{FAST}, \textit{F-FAST} for a fixed user set and \textit{D-FAST} for a dynamic user set.

\subsection{Fairness Assured Service Recommendation Strategy for A Fixed User Set - \textit{F-FAST}}

In order for users to reach a fair state as soon as possible, users with lower \textit{Top-N Fairness} should get more opportunities than users with higher \textit{Top-N Fairness}. Under the premise of limited service capacity, we preferentially meet the needs of recommendation list for users with lower \textit{Top-N Fairness}. To maintain a high level of recommendation list quality, when adjusting the original recommendation list $L$, we choose not to change the order of services in the list, and only delete from the list several services with insufficient capacity, and keep higher ranked services. Based on the above ideas, we design a heuristic algorithm based on greedy ideas.

The algorithm works as follows. Users are sorted according to their \textit{Top-N Fairness} from the lowest to the highest. The user with the lowest \textit{Top-N Fairness} will be recommended first. For a user $u_i$, a service with the highest rating score in $l(N)_i$ will be recommended as long as its capacity is still sufficient. If a service reaches its capacity constraint, the next best service in $l(N)_i$ will be recommended. After a service is recommended, the user's \textit{Top-N Fairness} and the capacity of the recommended service are updated. Then all users' \textit{Top-N Fairness} are sorted from the lowest to the highest again, and the next service is recommended in turn. This process ends when an attempt has been made to recommend every service in all users’ $l(N)_i$ (regardless of whether they are actually recommended to the user or not) or the capacities of all services have been exhausted. Finally, the algorithm fills the remaining empty positions in each user's recommendation list with services whose positions are larger than \textit{N} in his original list $l_i$ in sequence. The pseudo-code of \textit{F-FAST} is shown in \textbf{Algorithm 1}.

\begin{algorithm}
	\label{alg1}
	\caption{Fairness Assured Service Recommendation Algorithm for A Fixed User Set}
	\begin{algorithmic}[1]
		\Require
		$N$: Parameter Top-$N$;\newline 
		$l_{1}$, $l_{2}$,..., $l_{n}$: Original recommendation list of $n$ users;\newline
		$l(N)_1$, $l(N)_2$,..., $l(N)_n$: Original top-$N$ recommendation list of $n$ users;\newline 
		$R$: Rating matrix;\newline
		$c_1$, $c_2$,..., $c_m$: Capacity constraints of $m$ services;\newline
		$U_1$, $U_2$,..., $U_m$: $U_j$ list of $m$ services;\newline
		$F_1^{T-1}$, $F_2^{T-1}$,..., $F_n^{T-1}$: \textit{Top-N Fairness} of $n$ users up to last recommendation. 
		\Ensure
		$l^T_1$, $l^T_2$,..., $l^T_n$: Recommendation list for $n$ users in $T^{th}$ round;
		\For{$time = 0 \to n \times N - 1$}
		\State Sort users according to $F_i^{T-1}$ from lowest to highest
		\State $rec\_user \gets \text{user with the lowest } F_{i}^{T-1}$ 
		\For{$s_j$ in $l(N)_{rec\_user}$}
		\If{$c_j > 0$}
		\State $\text{insert }s_j\text{ into } l^T_{rec\_user}$;
		\State $c_j = c_j - 1$;
		\State $\text{Update }F_{rec\_user}^{T-1}$;
		\State break; 
		\EndIf
		\EndFor
		\EndFor
		\State Fill $l_i^T$ whose positions are larger than $N$ in $l_i$ in sequence;
		\State\Return $l^T_1$, $l^T_2$,..., $l^T_n$; 
	\end{algorithmic}	
\end{algorithm}

\textit{F-FAST} also has three properties, and for the relevant proofs of \textbf{THEOREM 1,2} refer to \textbf{Appendix \ref{Properties of F-FAST}}. 

\begin{theorem}
	The sum of Top-$N$ Fairness of all the users in each round is equal to zero, $\sum_{u_i\in U}F_i^T = 0$.
\end{theorem} 

\begin{theorem}
	Variance among Top-$N$ fairness of all users $D(F_i^T)$ converges to 0 with the recommended round $T$.
\end{theorem}

\begin{theorem}
	The system can reach the Individual Level Fair Status, $\forall u_i\in U, F_i^T = 0$.
\end{theorem}

\textbf{Theorem 1} indicates that the sum of \textit{Top-N Fairness} is stable, and \textbf{Theorem 2} indicates  \textit{F-FAST} can ensure long-term fairness for multi-round recommendations. By combining \textbf{Theorem 1} and \textbf{Theorem 2}, we can conclude that $F_i^T$ of each user in the system will eventually converge to 0, so that the system can reach the \textit{Individual Level Fair Status} which is \textbf{Theorem 3}.

\subsection{Fairness Assured Service Recommendation Strategy for A Dynamic User Set - \textit{D-FAST}}\label{Fair Recommendation Algorithm for Dynamic User Sets}
\textit{D-FAST} is applied to the situation where the user set receiving recommendations changes from time to time. In this situation, we cannot guarantee the validity of \textbf{Theorem 1}. The average fairness of users receiving recommendations is different in each round, which leads to changes in the baseline of \textit{Top-N Fairness}. 

Therefore, before generating the recommendation lists, the user's \textit{Top-N Fairness} needs to be calculated again to make up for the baseline change. The strategy works as follows: the average \textit{Top-N Fairness} of users is recorded after each round, and at the beginning of a new round, the user's \textit{Top-N Fairness} is updated by adding the difference between the average \textit{Top-N Fairness} in his last round and the average \textit{Top-N Fairness} in this round. In addition, for a new user, his \textit{Top-N Fairness} and the average \textit{Top-N Fairness} of the last round will both be set to zero. The remaining operations are the same as \textbf{Algorithm 1}.

\subsection{Time Complexity}
The time complexity of \textit{F-FAST} is analyzed as follows. When recommending a service, \textit{F-FAST} first sorts users according to their \textit{Top-N Fairness}. The complexity of sorting $n$ users will be $O(nlog(n))$ when using the Quick Sort Algorithm or the Merge Sort Algorithm. Then \textit{F-FAST} recommends a service to the user with the lowest \textit{Top-N Fairness}. These are operations with a single instruction, so the complexity of recommending an item is $O(nlog(n) + 1)$. In a round of recommendations, we need to recommend a maximum of $n \times N$ services, where \textit{N} is a small constant. So, in a round of recommendations, the worst case time complexity of \textit{F-FAST} is $O(n^2log(n))$. 

There is only one additional step for \textit{D-FAST} which is at the beginning of each round, that is, to update the \textit{Top-N Fairness} of all users, and its time complexity is also $O(n^2log(n))$.

\section{Experiments}\label{Experiments}
\textbf{Datasets and Metrics}  We conduct experiments on a real-world dataset and a synthetic datasets. Our code and datasets are released on Zenodo \footnote{https://zenodo.org/record/3661863\#.XkJGb2gzZPY}.

\textbf{Yelp dataset} The data is provided by the Yelp Dataset Challenge \cite{challenge2019yelp}.  We  select  two  cities  with  the  largest  number  of  businesses,  i.e.,  Phoenix  and Toronto. After filtering out users less than 10 reviews and businesses less than 30 reviews, we obtain the dataset for Phoenix, which contains 11,252 users, 3774 businesses and 194,188 reviews. The Toronto dataset contains 8867 users, 3,505 businesses, and 1,190,64 reviews.

\textbf{Synthetic datasets} We generate synthetic datasets to test the performance of the algorithms under different parameter settings. For this purpose, we generate 4 synthetic datasets with different situations of capacity conflicts when \textit{N} is set to 5:
\begin{itemize}
	\item \textit{Very Popular Services}: the number of users in $U_j$  is more than 2 times its capacity.
	\item \textit{Popular Services}: the number of users in $U_j$  is 1-2 times its capacity.
	\item \textit{Ordinary Services}: the number of users in $U_j$  is 0.9-1.0 times its capacity.
	\item \textit{Unpopular Services}: the number of users in $U_j$ is 0.9 times its capacity.
\end{itemize}
The capacity of each service is a random number from 50 to 100. Each dataset contains 800 users  and  50  services.

\textbf{Metrics} We measure the total quality of the recommendations of each user and the variance of the \textit{Top-N Fairness} of all users at the same time.

\subsection{Compared Approaches}
This is the first time that the fairness assured multi-round recommendation problem for services with capacity constraints is defined and there is no existing algorithm for this problem. Thus, we compare our approach against the following three baseline methods.

\textbf{Integer Linear Programming} We use Integer Linear Programming (ILP) to maximize the quality of recommendations. We take the capacity constraints as the limitations and the quality of recommendations as the target. 

\textbf{Greedy Algorithm to maximize quality of recommendation}  The size of the problem that can be solved by ILP is limited, so when processing the Yelp dataset, we replace the ILP method with a greedy heuristic algorithm. The idea of the algorithm is to recommend services that ensure the best quality of recommendation each time as long as the capacity constraint of a service is not violated.

\textbf{Random Strategy} For a service $s_j$, we randomly select a number of users which equals $s_j$'s capacity constraint from its $U_j$ list in each round. Obviously, this strategy can also ensure the \textit{Top-N Fairness} in the long run.

\subsection{Results on Yelp Dataset}\label{Results on Yelp Dataset}
\begin{figure}[!h]
	\centering
	\subfigure[Phoenix]{
		\begin{minipage}[t]{0.2\linewidth}
			\centering
			\includegraphics[width=\textwidth,height=2cm]{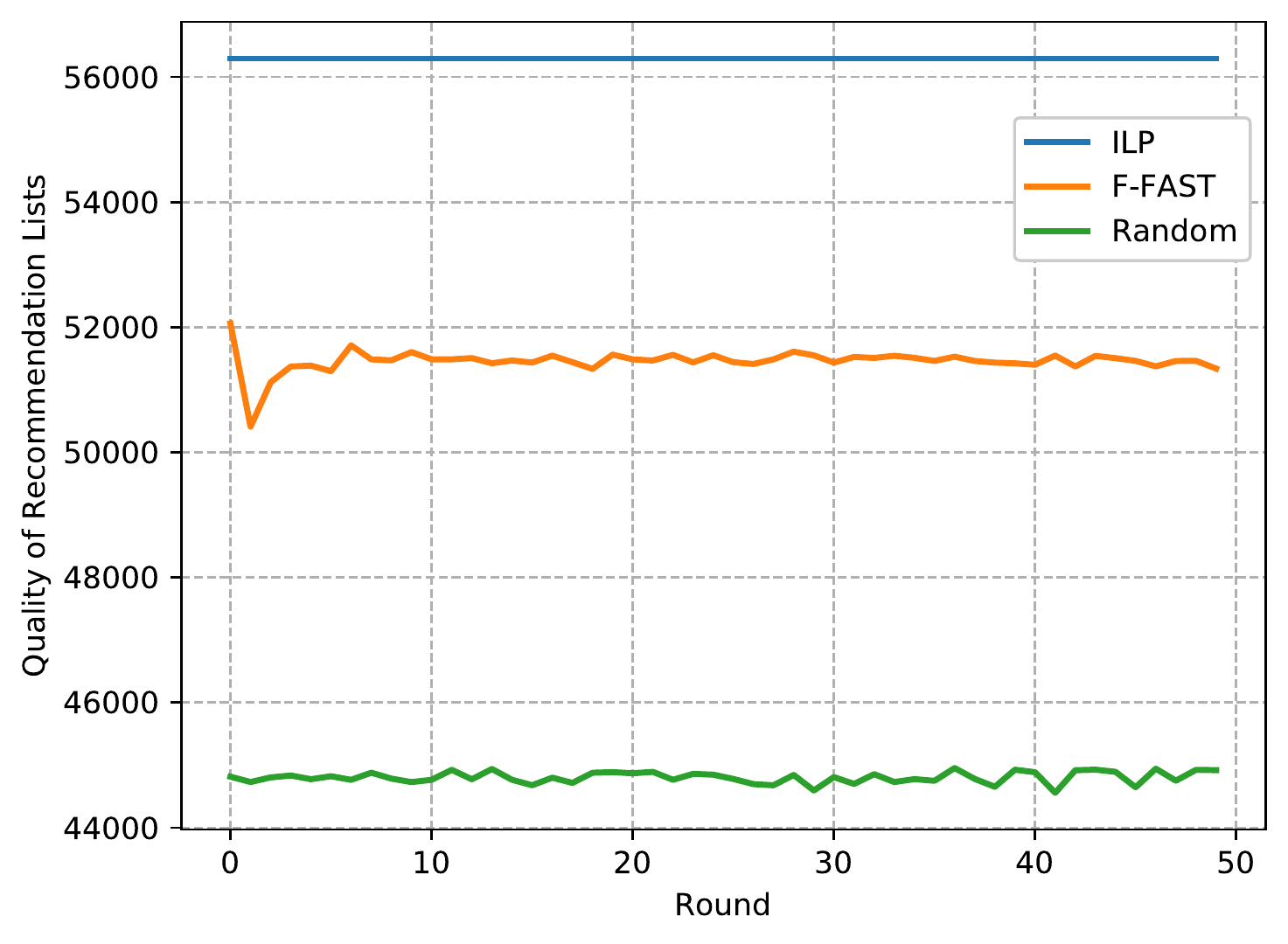}
		\end{minipage}%
	}%
	\quad
	\subfigure[Phoenix]{
		\begin{minipage}[t]{0.2\linewidth}
			\centering
			\includegraphics[width=\textwidth,height=2cm]{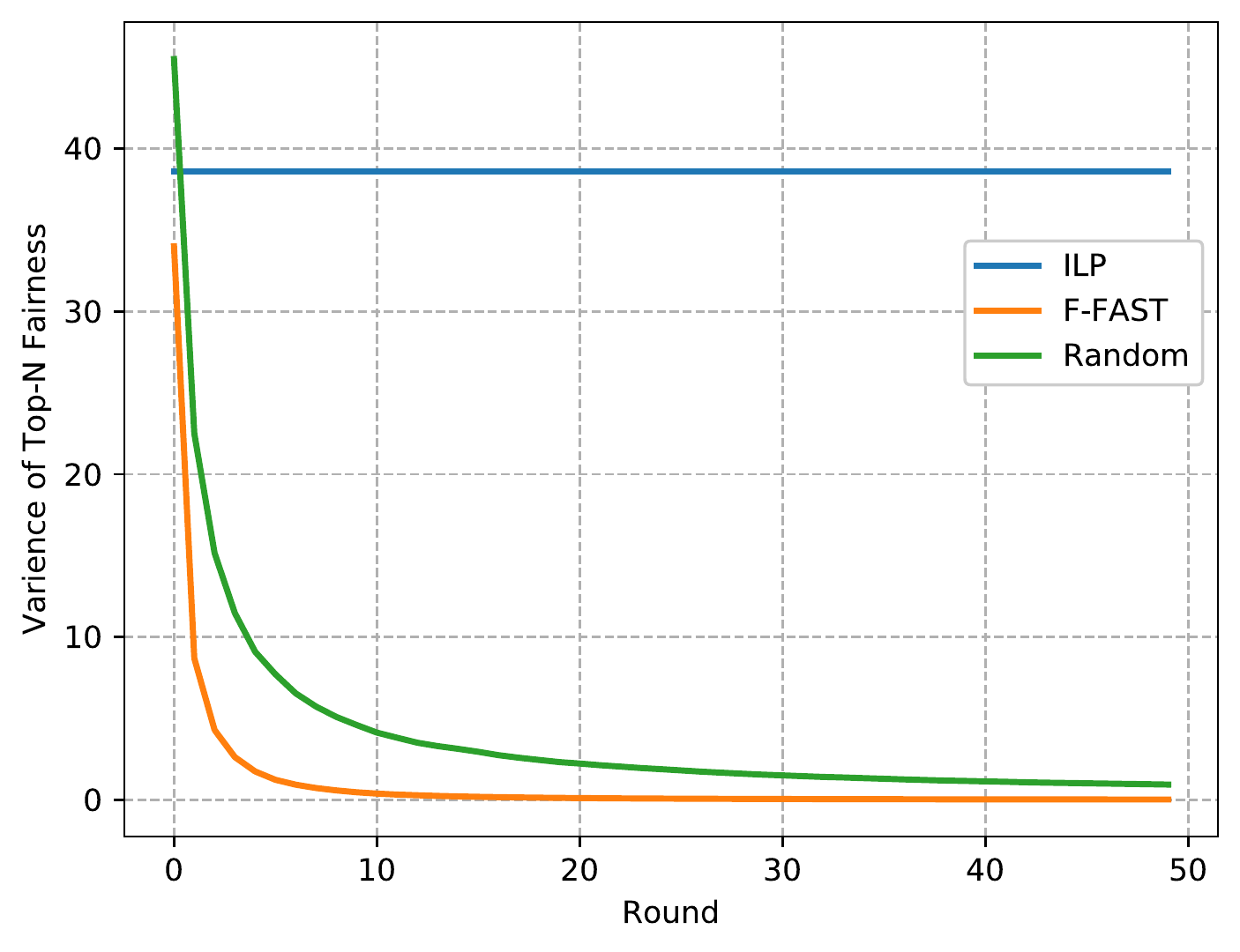}
		\end{minipage}%
	}%
	\quad
	\subfigure[Toronto]{
		\begin{minipage}[t]{0.2\linewidth}
			\centering
			\includegraphics[width=\textwidth,height=2cm]{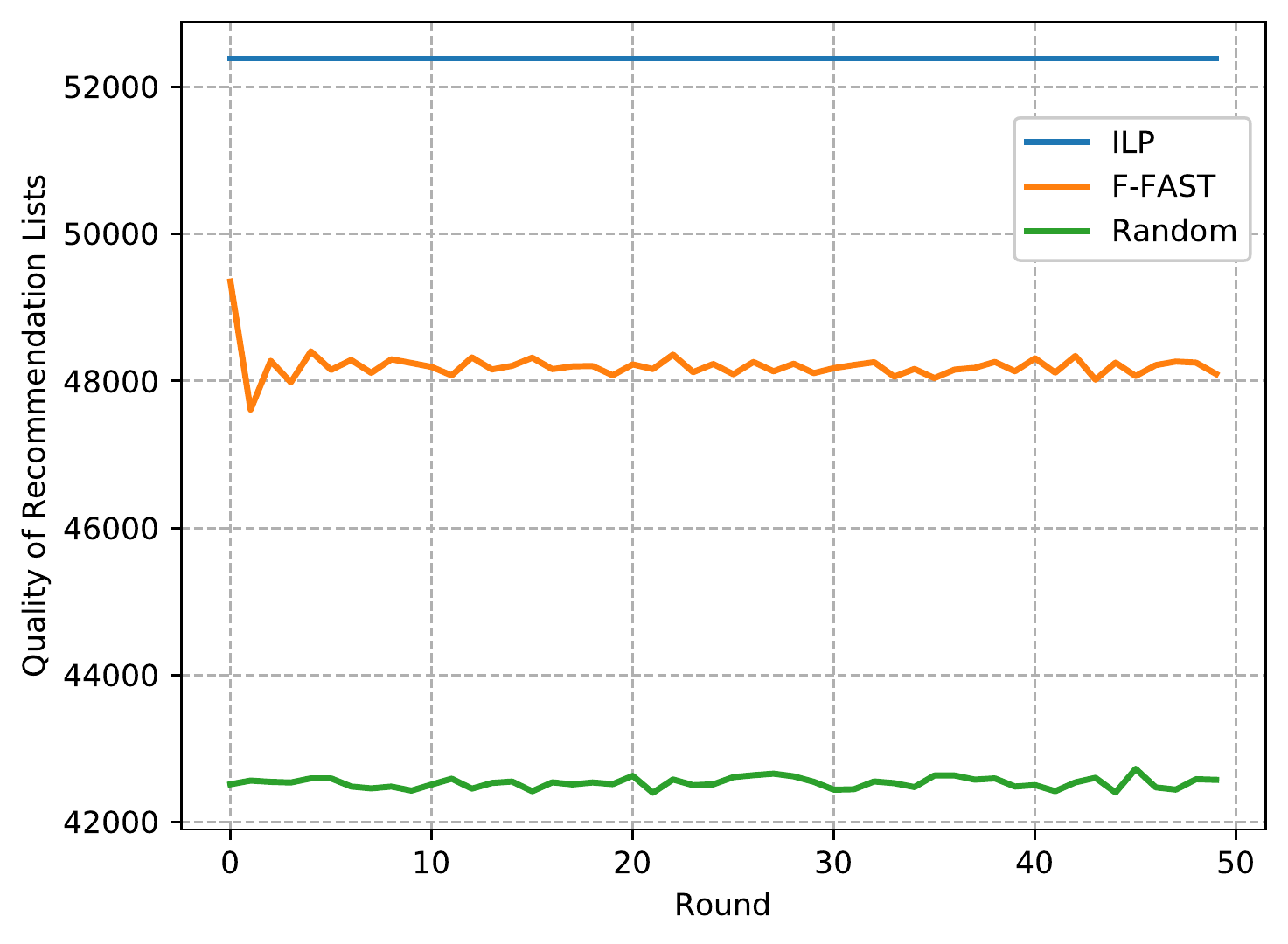}
		\end{minipage}%
	}%
	\quad
	\subfigure[Toronto]{
		\begin{minipage}[t]{0.2\linewidth}
			\centering
			\includegraphics[width=\textwidth,height=2cm]{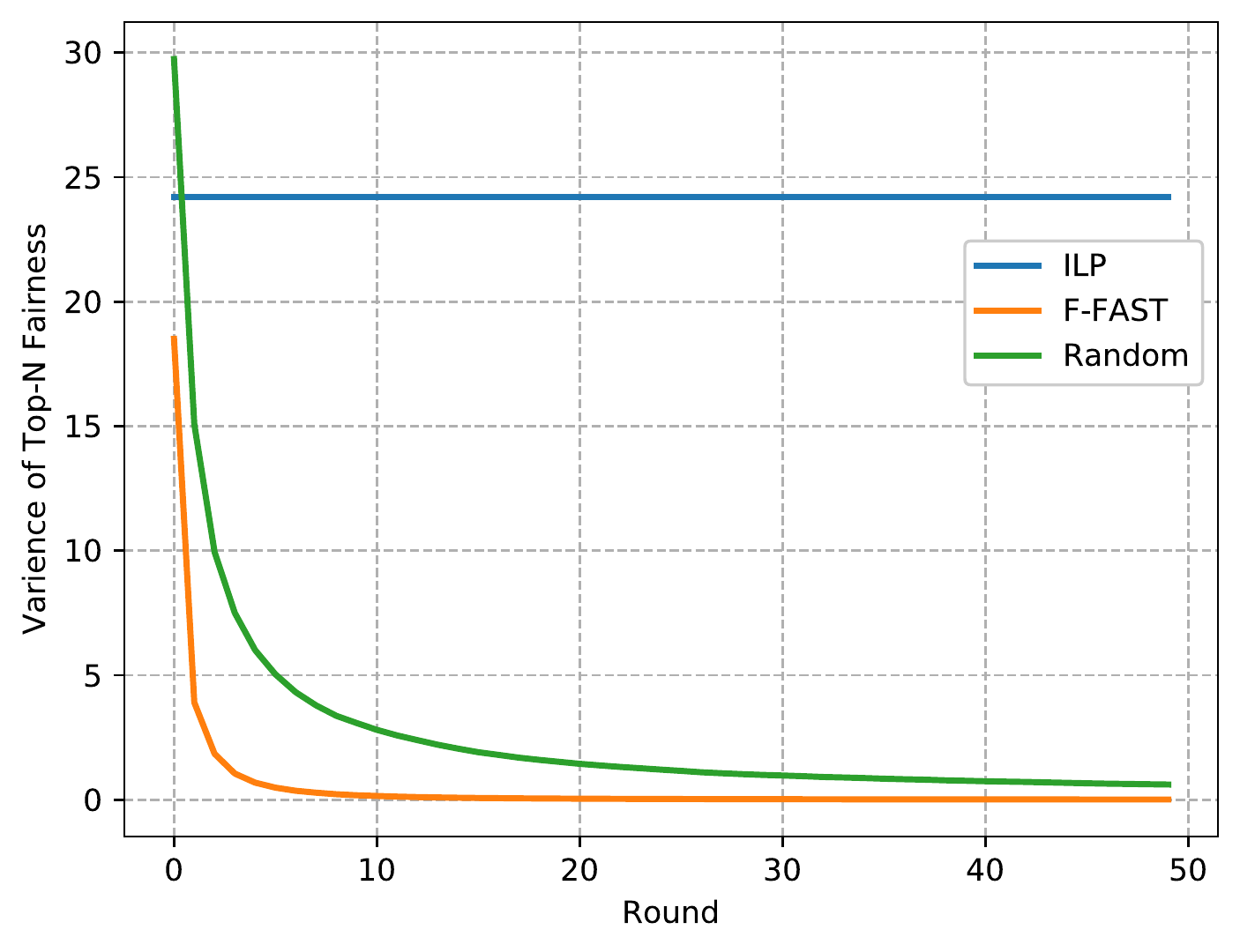}
		\end{minipage}%
	}%
	\centering
	\caption{Quality of recommendation and variance of \textit{Top-N Fairness} on Yelp Dataset}
	\label{fig2}
\end{figure}

We perform 50 rounds of recommendations on a fixed user set on the Yelp dataset. Since recommendations are pushed to a fixed user set, \textit{F-FAST} is executed to generate the recommendations. In this experiment, we set \textit{N} to 5, and Figure \ref{fig2} shows the results.

From the figure, we can draw two conclusions. First, \textit{F-FAST} makes the system reach a relatively balanced state (variance of \textit{Top-N Fairness} approaches zero) faster, and the degree of fairness is also the highest. Although the random strategy can also achieve a relatively fair situation, compared to \textit{F-FAST}, the speed at which it arrives at a stable status of fairness is much slower. Second, regarding recommendation quality, \textit{F-FAST} has a small loss while the random strategy leads to significant losses. \textit{F-FAST} loses $7\%$ of recommendation quality compared with the ILP method but it is nearly $20\%$ higher than the random strategy. 

\subsection{Results on Synthetic Datasets}

\subsubsection{Comparisons between Different Levels of Capacity Constraints}\label{Comparison between Different Levels of Capacity Conflicts}
\begin{figure*}[!h]
	\centering
	\subfigure[Dataset 1]{
		\begin{minipage}[t]{0.2\linewidth}
			\centering
			\includegraphics[width=\textwidth,height=2cm]{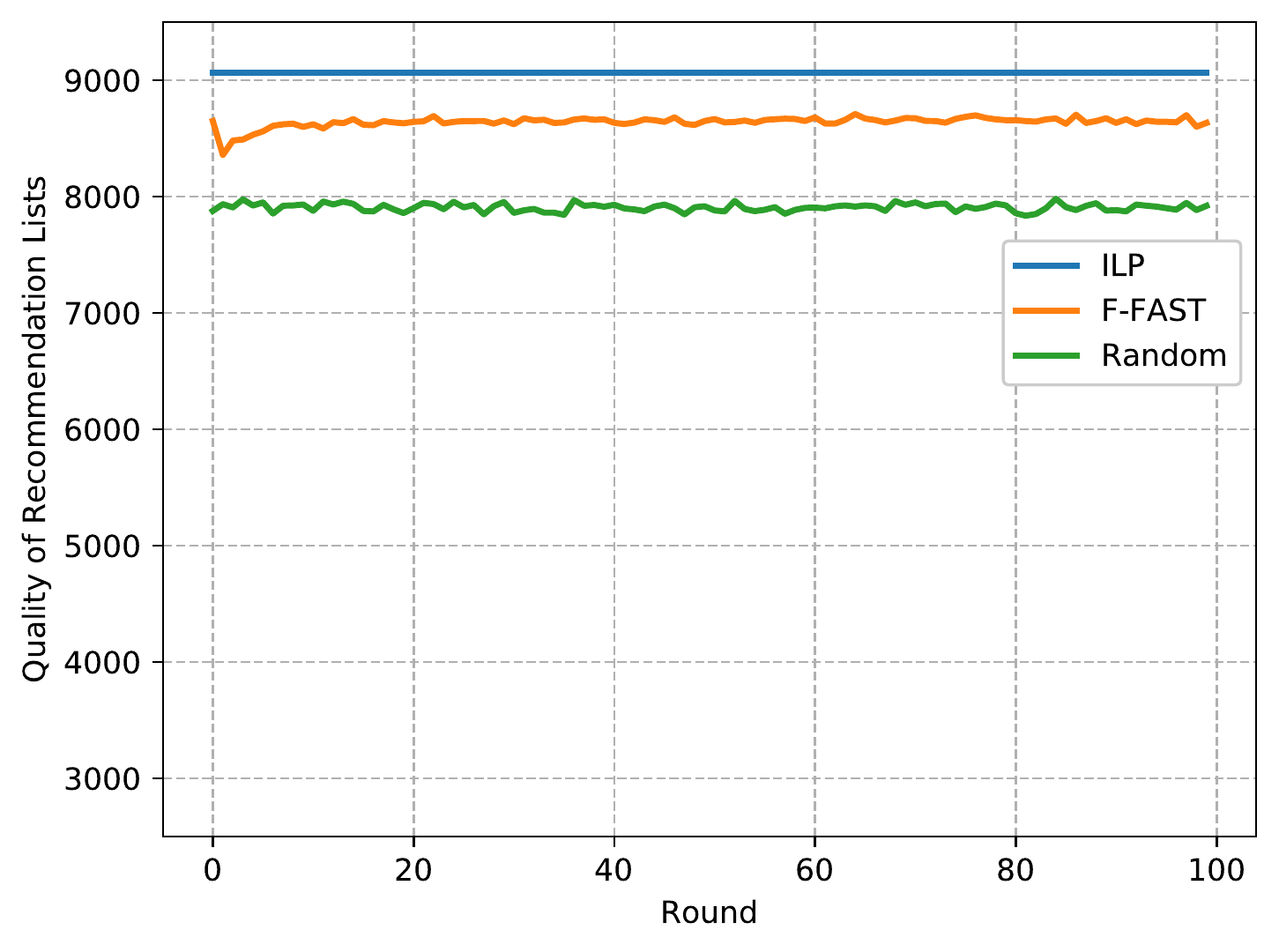}
		\end{minipage}%
	}%
	\subfigure[Dataset 2]{
		\begin{minipage}[t]{0.2\linewidth}
			\centering
			\includegraphics[width=\textwidth,height=2cm]{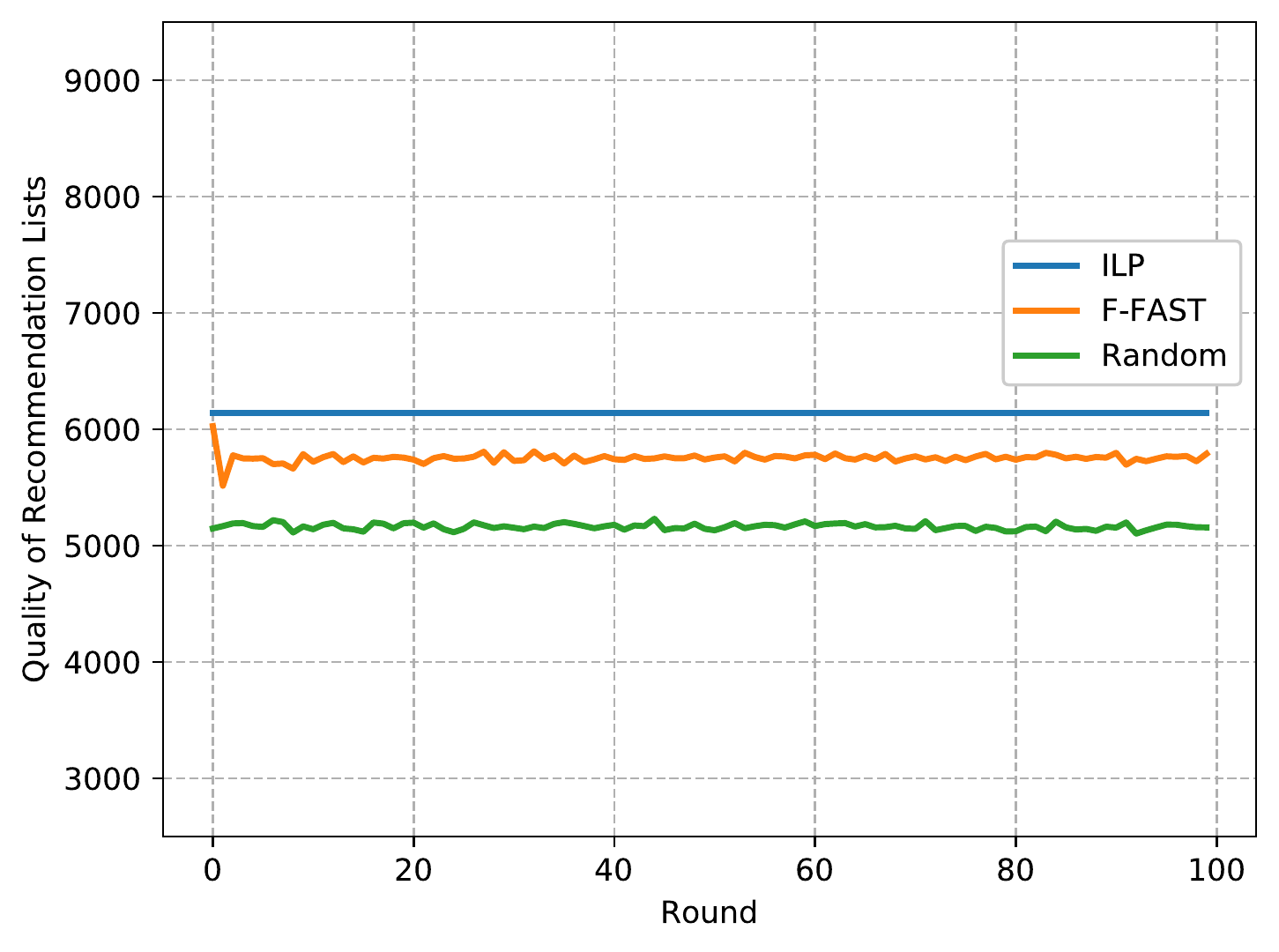}
		\end{minipage}%
	}%
	\subfigure[Dataset 3]{
		\begin{minipage}[t]{0.2\linewidth}
			\centering
			\includegraphics[width=\textwidth,height=2cm]{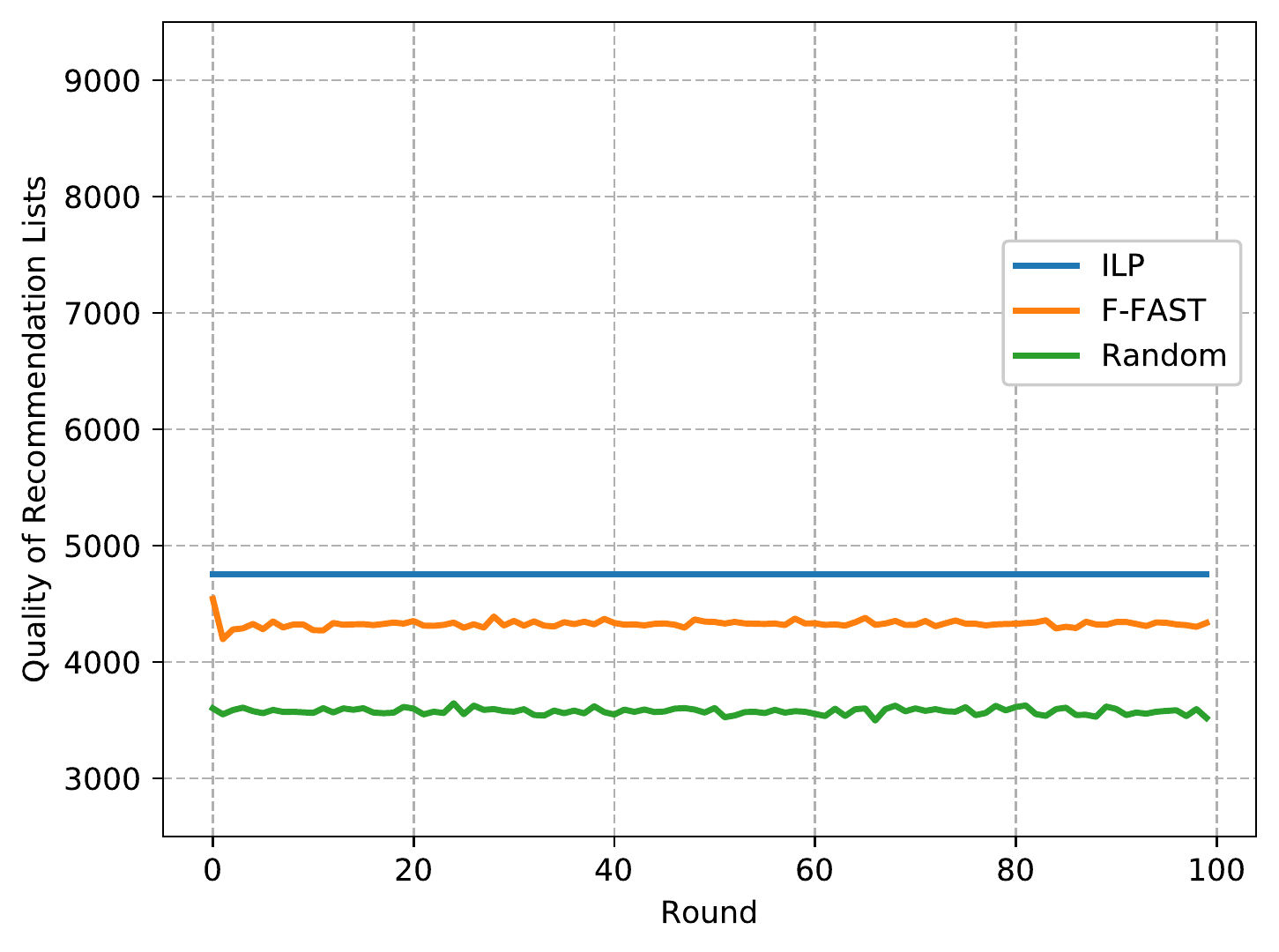}
		\end{minipage}
	}%
	\subfigure[Dataset 4]{
		\begin{minipage}[t]{0.2\linewidth}
			\centering
			\includegraphics[width=\textwidth,height=2cm]{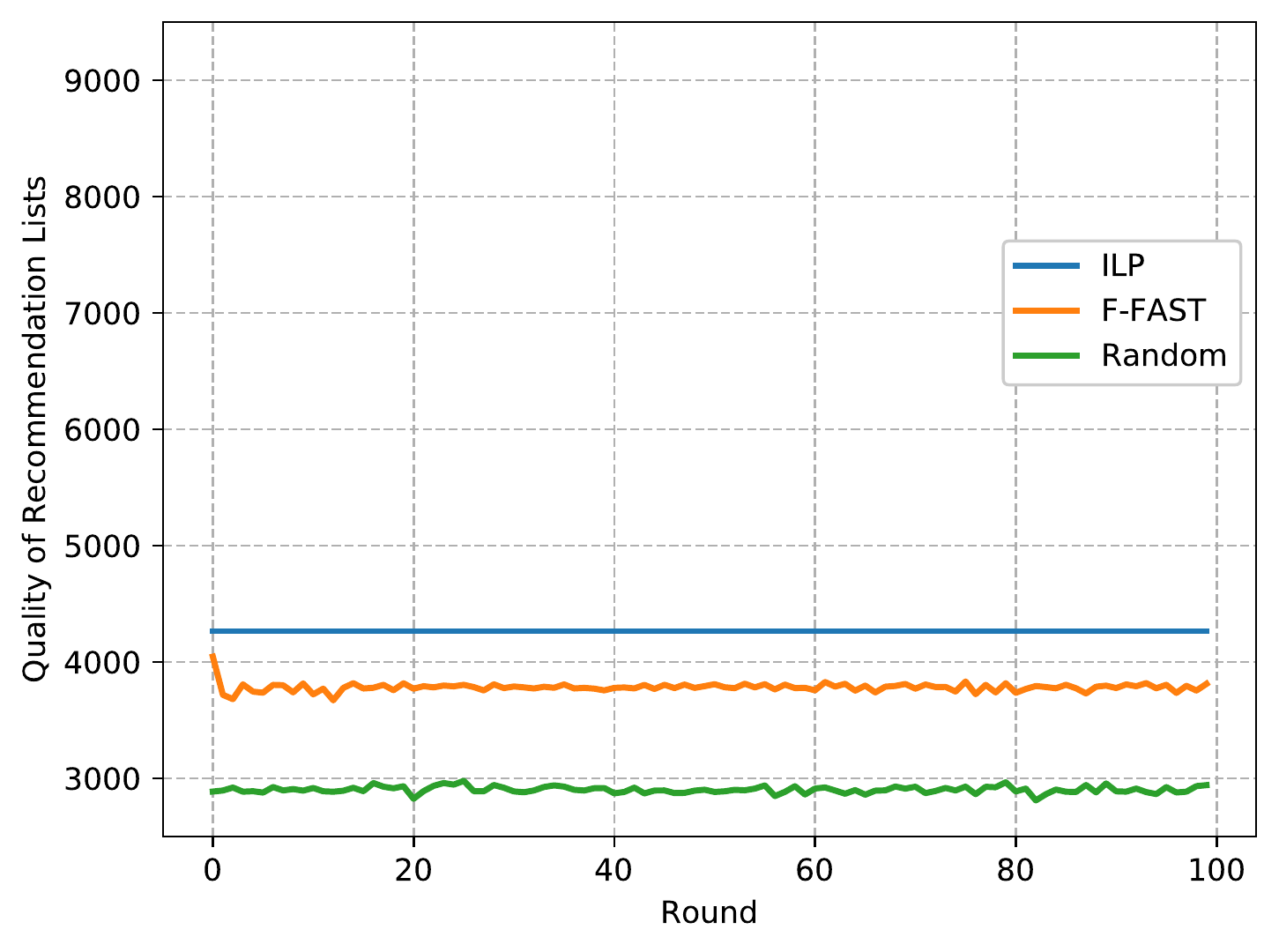}
		\end{minipage}
	}%
    \quad
    	\subfigure[Dataset 1]{
    	\begin{minipage}[t]{0.2\linewidth}
    		\centering
    		\includegraphics[width=\textwidth,height=2cm]{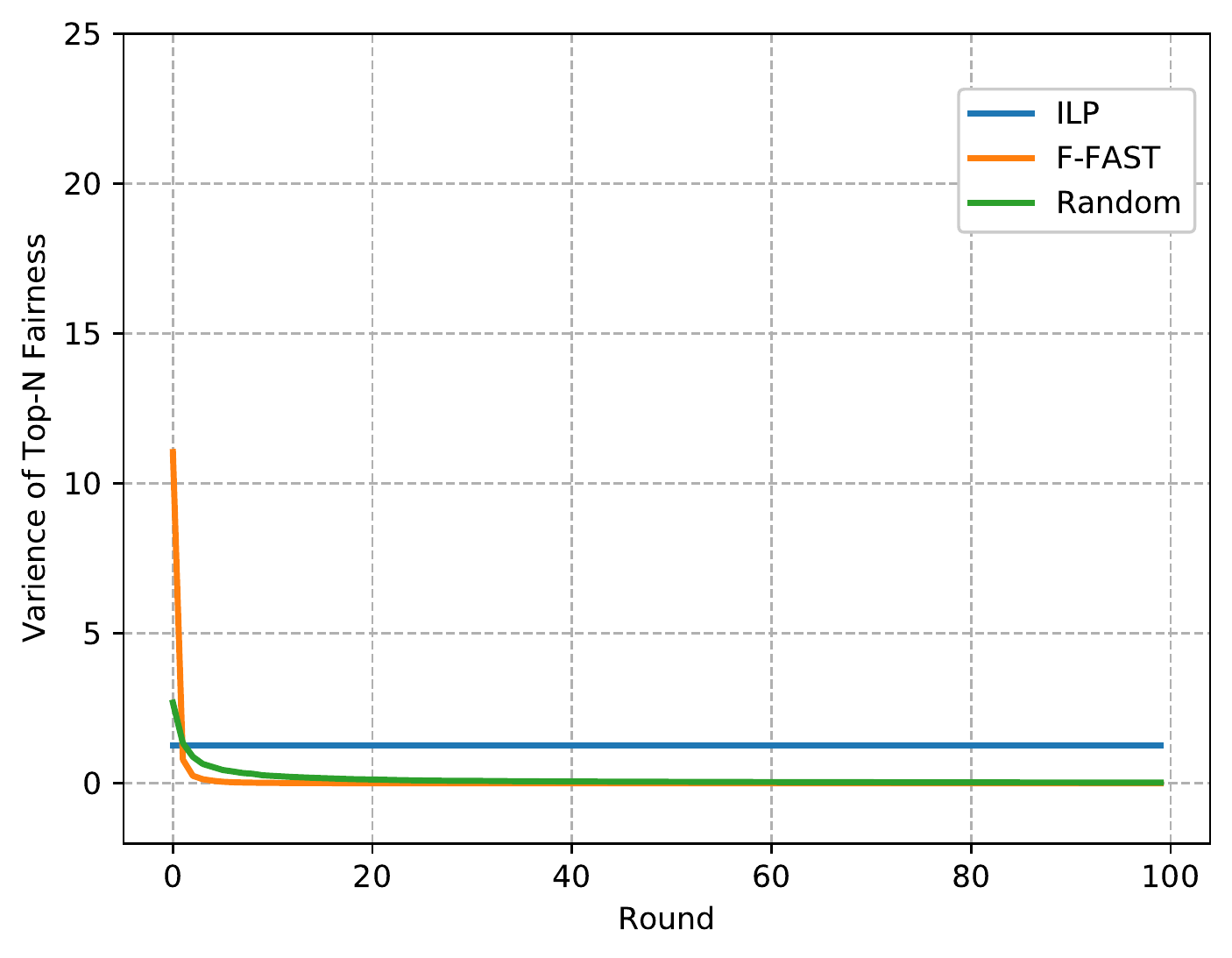}
    	\end{minipage}%
    }%
    \subfigure[Dataset 2]{
    	\begin{minipage}[t]{0.2\linewidth}
    		\centering
    		\includegraphics[width=\textwidth,height=2cm]{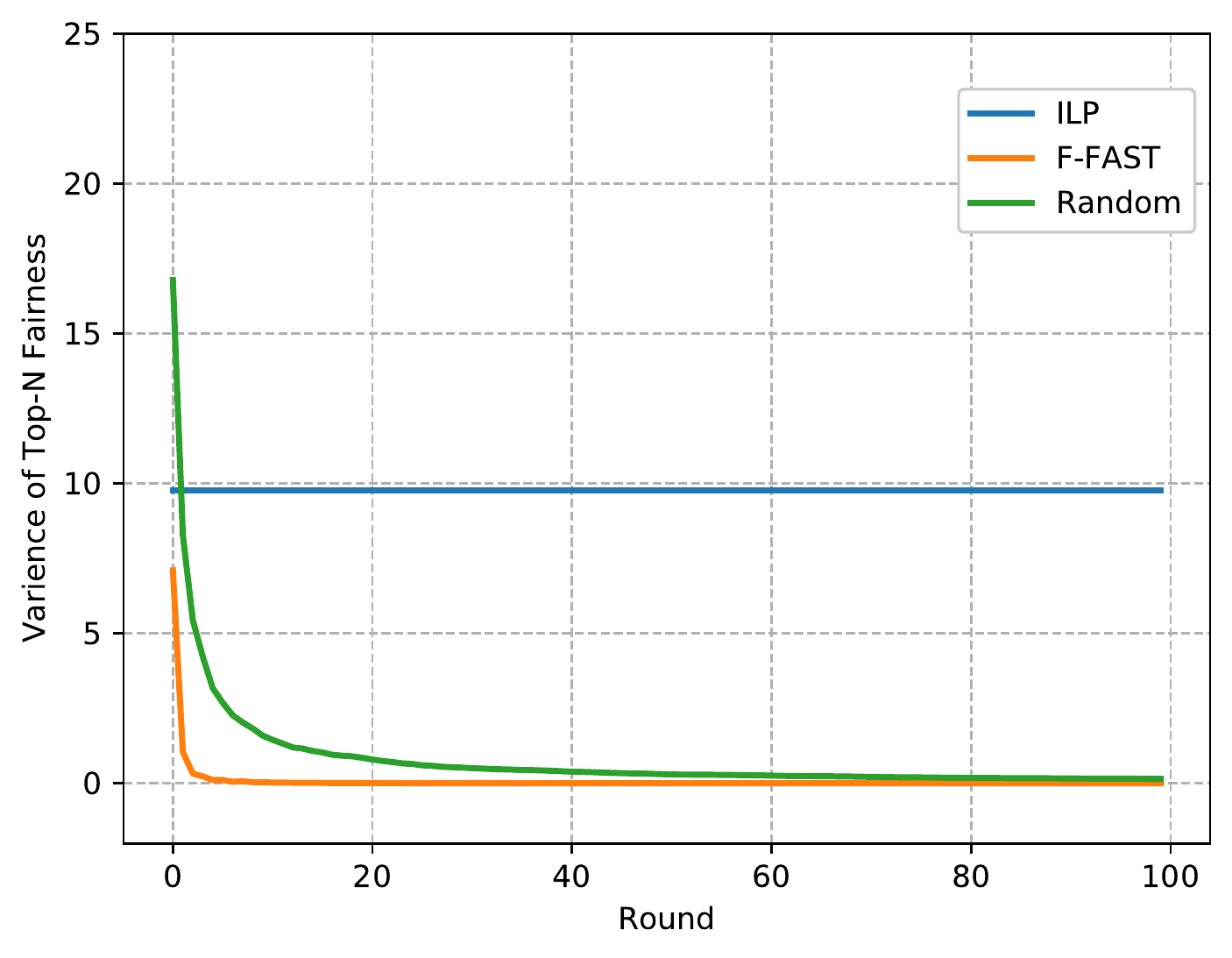}
    	\end{minipage}%
    }%
    \subfigure[Dataset 3]{
    	\begin{minipage}[t]{0.2\linewidth}
    		\centering
    		\includegraphics[width=\textwidth,height=2cm]{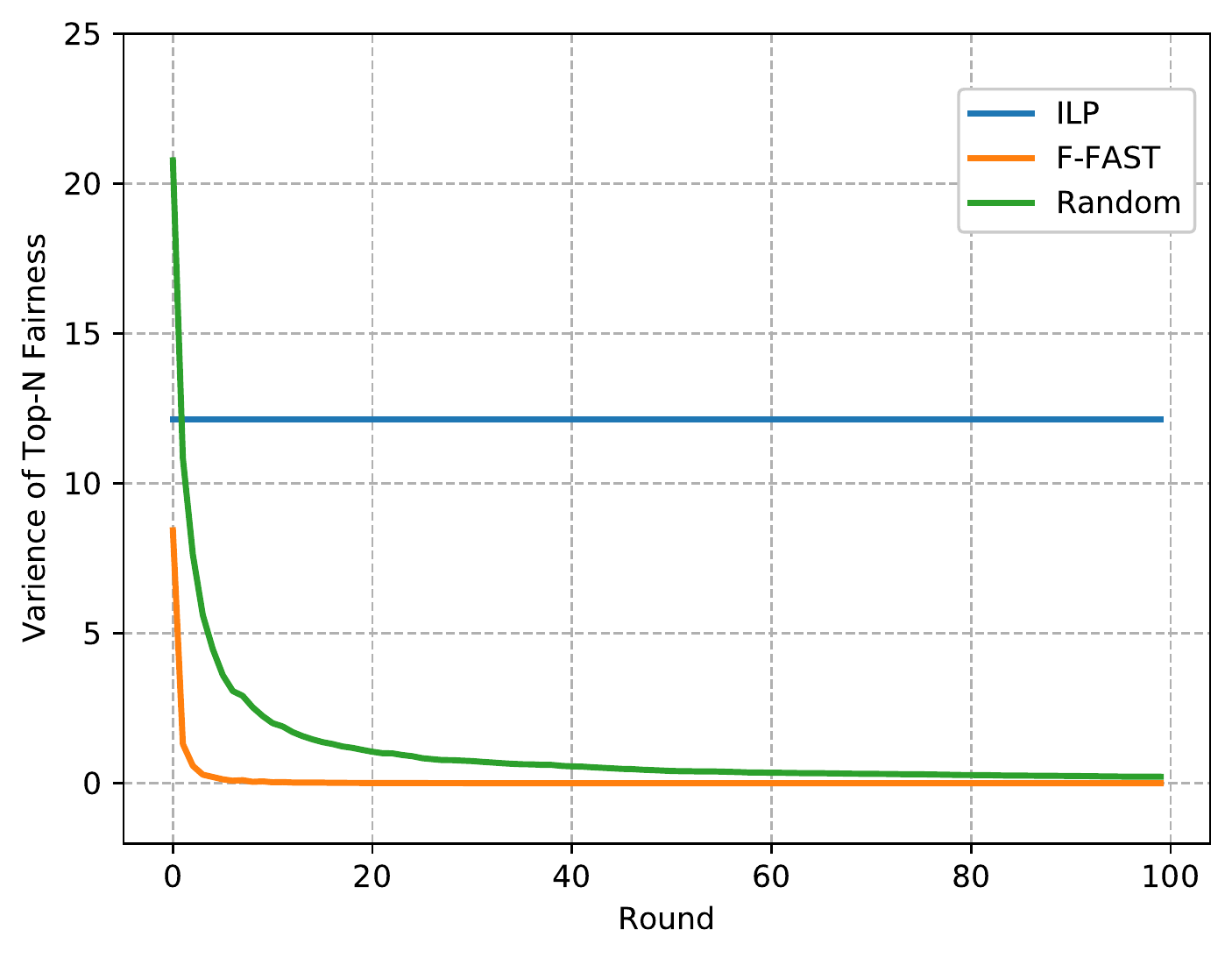}
    	\end{minipage}
    }%
    \subfigure[Dataset 4]{
    	\begin{minipage}[t]{0.2\linewidth}
    		\centering
    		\includegraphics[width=\textwidth,height=2cm]{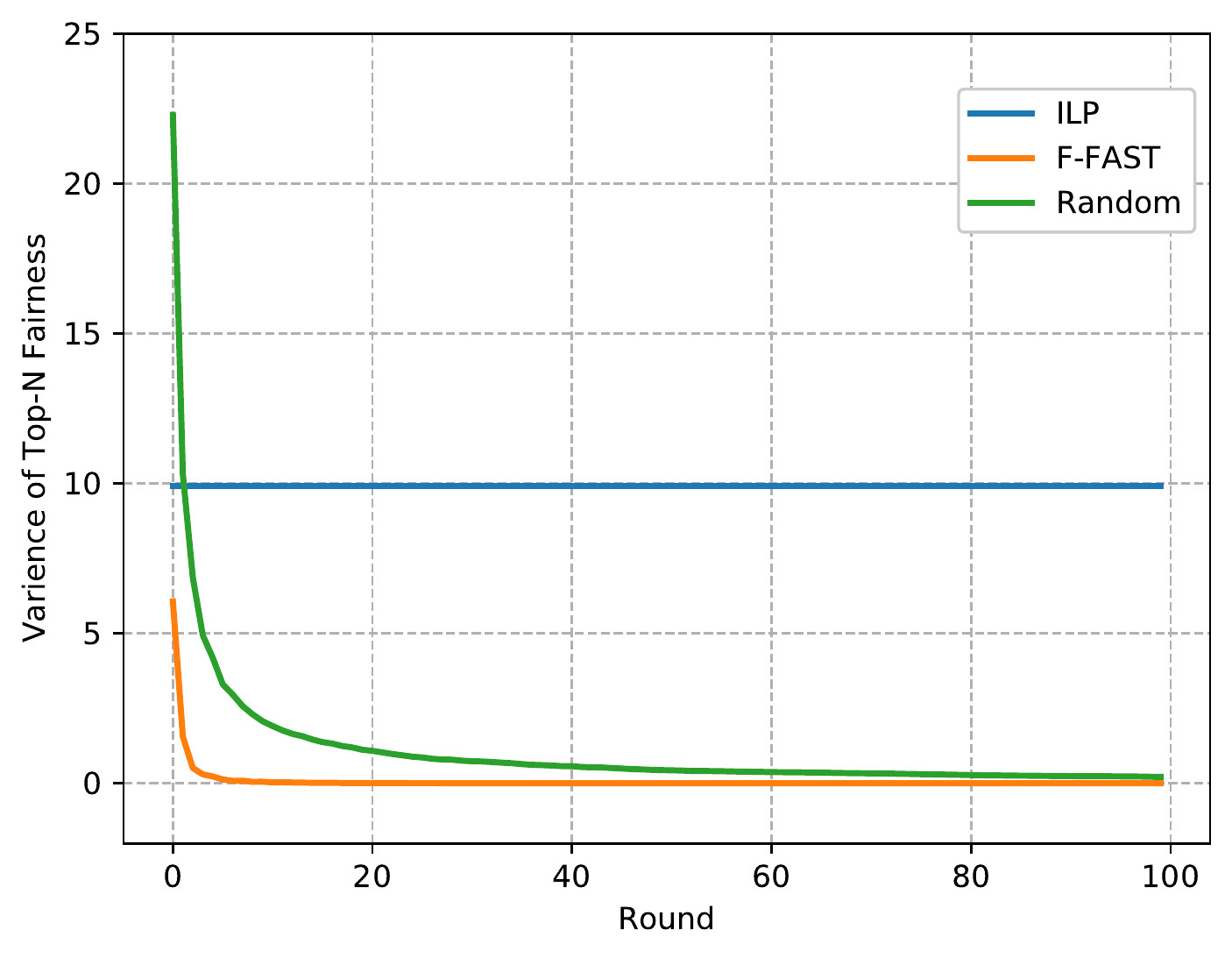}
    	\end{minipage}
    }%
	\centering
	\caption{Recommendation quality and variance of \textit{Top-N Fairness} under Different Levels of Capacity Constraints}
	\label{fig3}
\end{figure*}

We conduct 100 rounds of recommendations on a fixed user set on four synthetic datasets. In four groups of experiments, $N$ is uniformly set to 5, and results are shown in Figure \ref{fig3}. It can be seen from Figures \ref{fig3}a, \ref{fig3}b, \ref{fig3}c and \ref{fig3}d that as capacity conflict being more intense, the quality of recommendations tends to decrease. The reason for this is when capacity conflict becomes more intense, users have less chances of being assigned one of the top-$N$ services in his original list, which in turn leads to a decrease in quality. Figures \ref{fig3}e, \ref{fig3}f, \ref{fig3}g and \ref{fig3}h show that as the capacity conflicts become more intense, the total fairness of ILP and the random strategy basically show a downward trend, while the \textit{F-FAST} is not affected by the intensity of capacity conflicts. At the same time, \textit{F-FAST} arrives at a stable status of fairness faster than the random strategy in all scenarios.  

\subsubsection{Comparisons between Different \textit{N}}\label{Comparison between Different Top-N}
\begin{figure*}[!h]
	\centering
	\subfigure[$N = 3$]{
		\begin{minipage}[t]{0.2\linewidth}
			\centering
			\includegraphics[width=\textwidth,height=2cm]{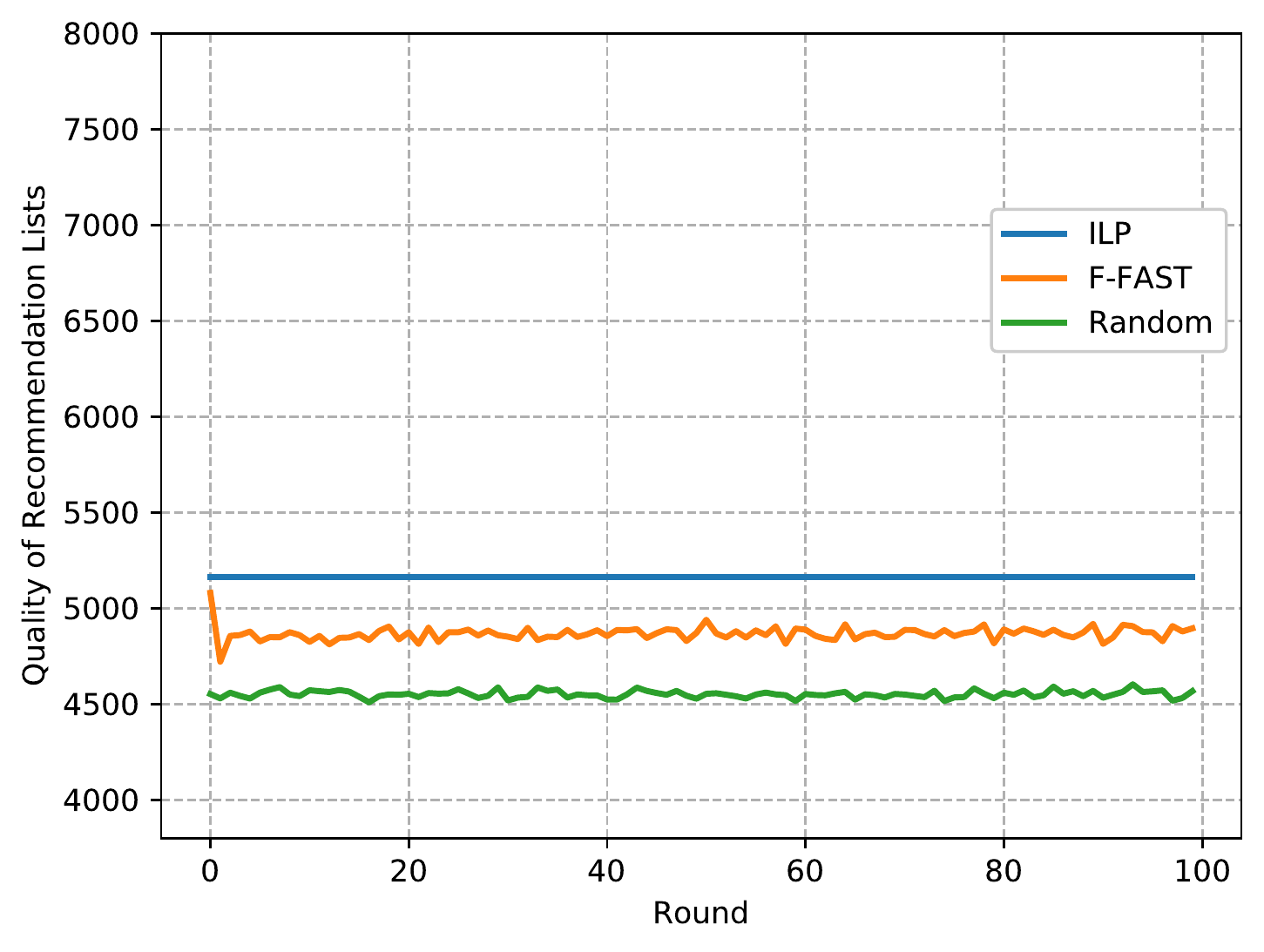}
		\end{minipage}%
	}%
	\subfigure[$N = 5$]{
		\begin{minipage}[t]{0.2\linewidth}
			\centering
			\includegraphics[width=\textwidth,height=2cm]{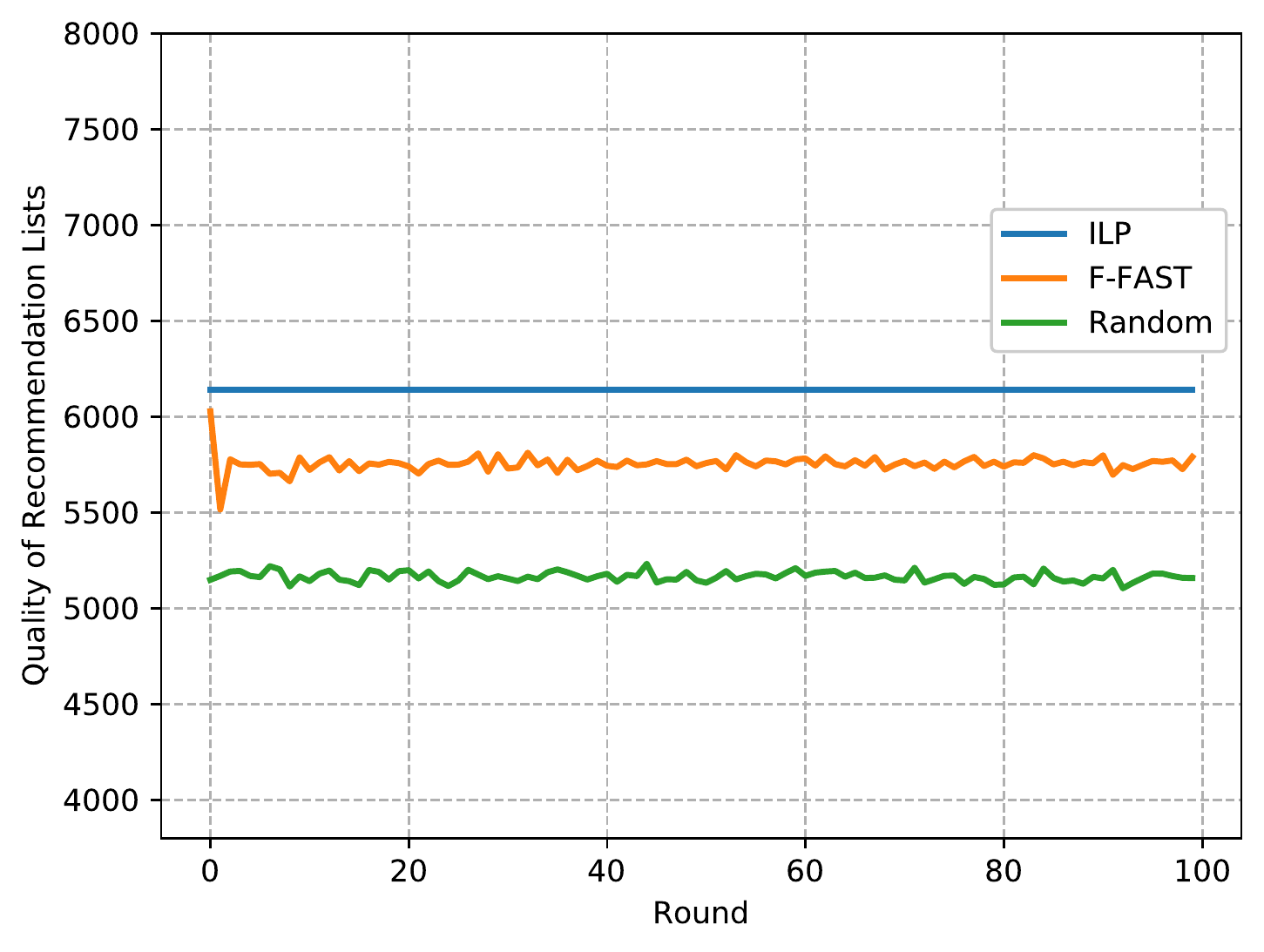}
		\end{minipage}%
	}%
	\subfigure[$N = 10$]{
		\begin{minipage}[t]{0.2\linewidth}
			\centering
			\includegraphics[width=\textwidth,height=2cm]{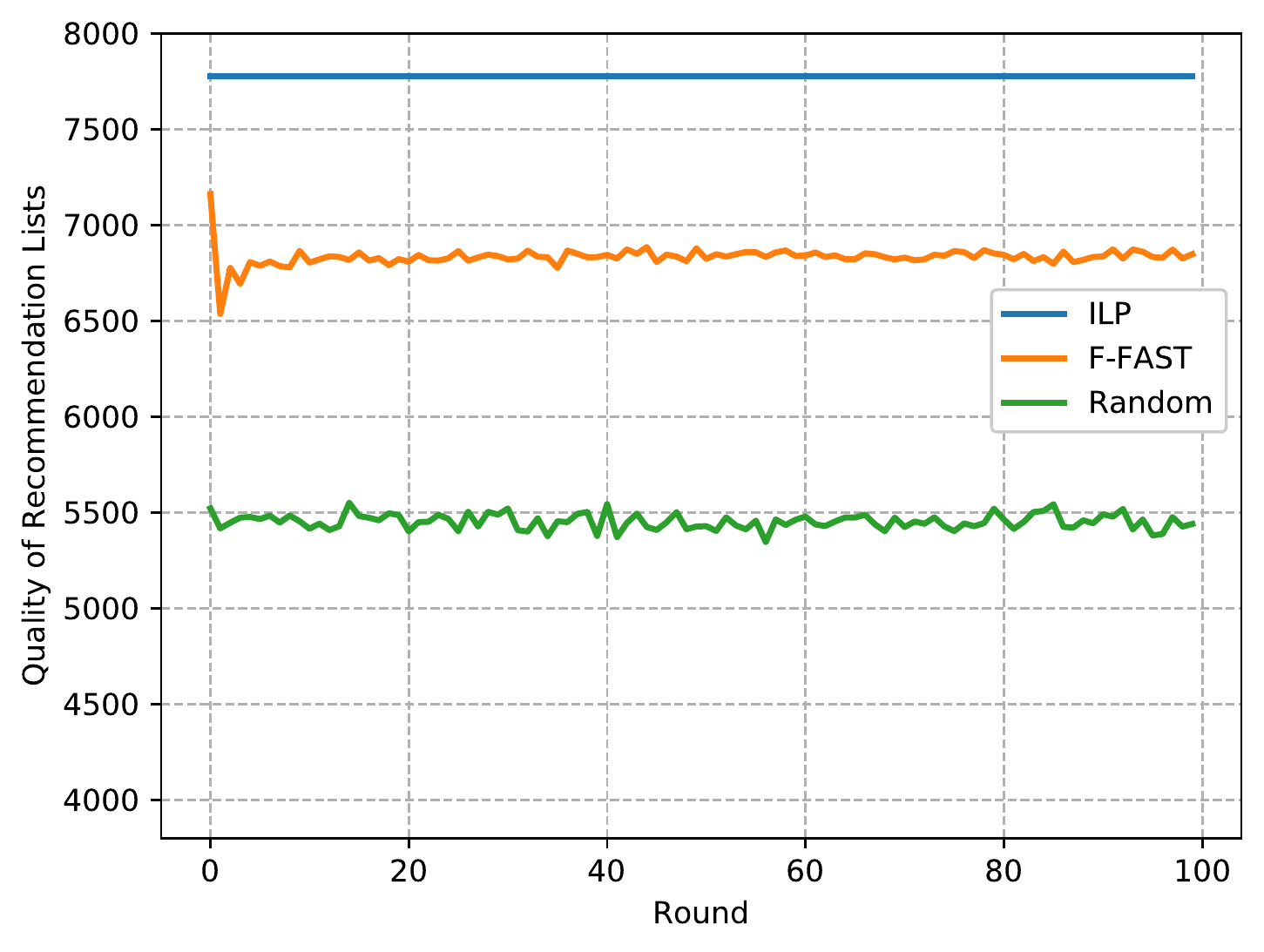}
		\end{minipage}%
	}%
	\subfigure[$N = 15$]{
		\begin{minipage}[t]{0.2\linewidth}
			\centering
			\includegraphics[width=\textwidth,height=2cm]{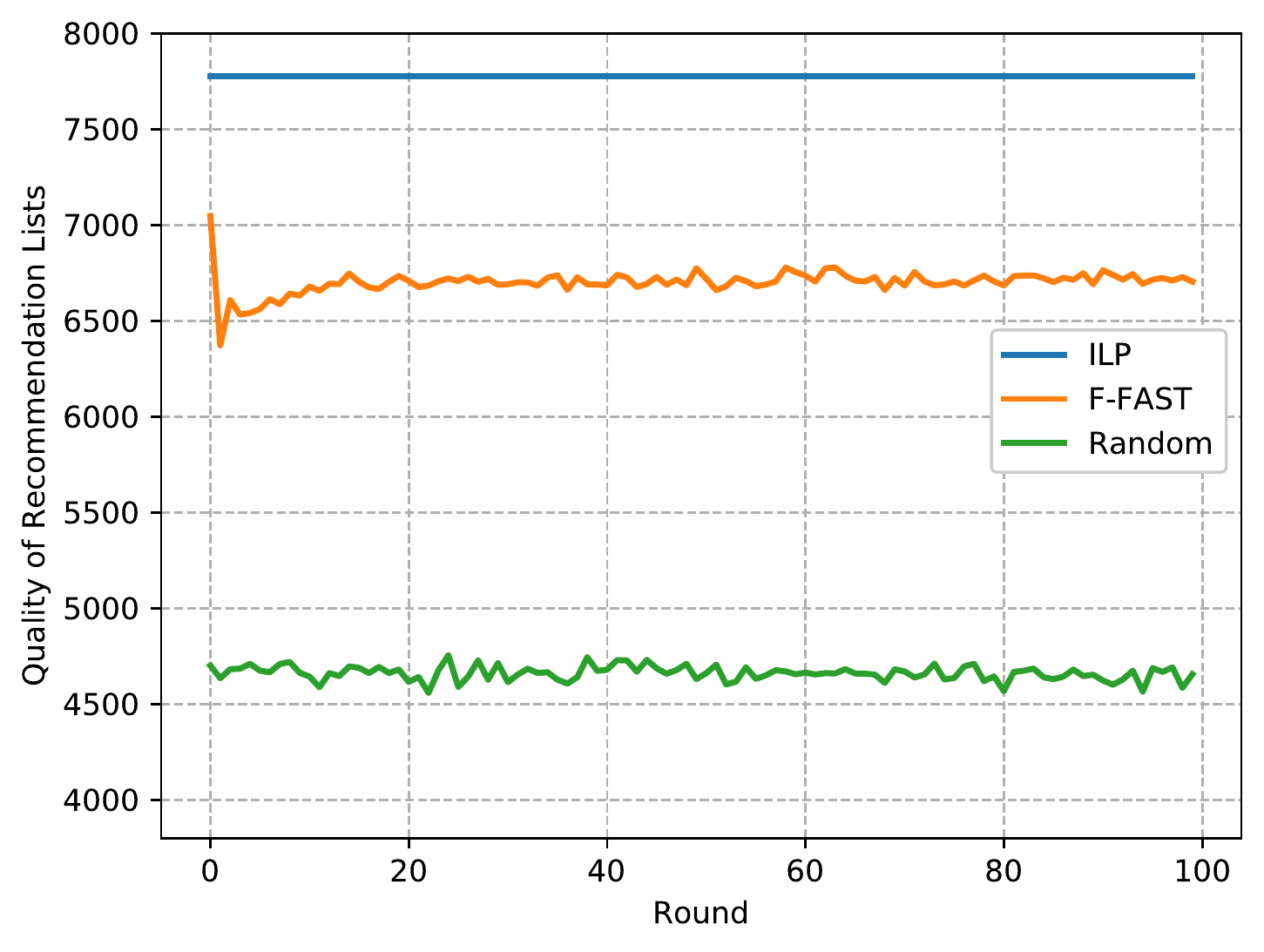}
		\end{minipage}%
	}%
	\subfigure[$N = 20$]{
		\begin{minipage}[t]{0.2\linewidth}
			\centering
			\includegraphics[width=\textwidth,height=2cm]{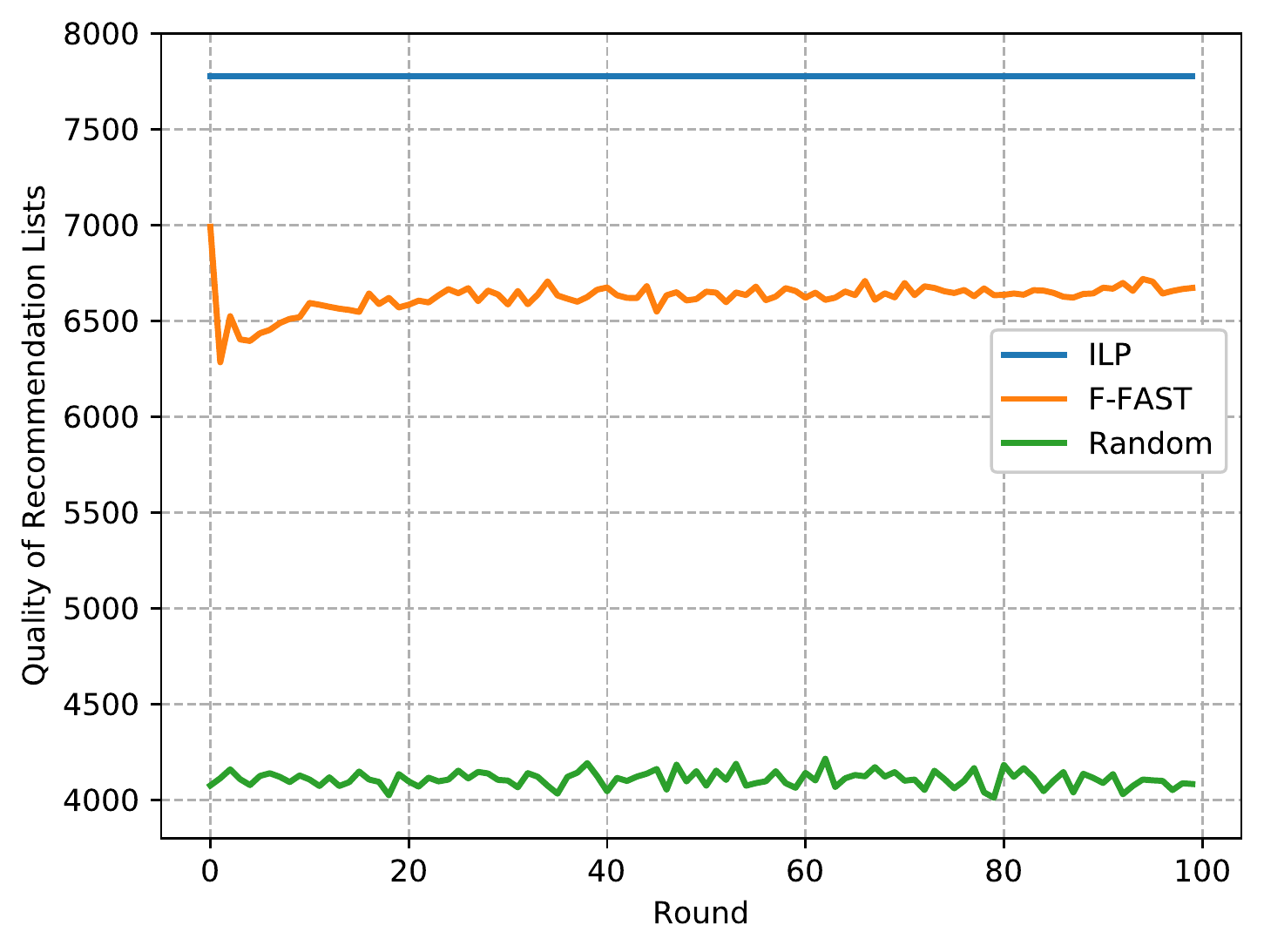}
		\end{minipage}%
	}%
    \quad
    	\subfigure[$N = 3$]{
    	\begin{minipage}[t]{0.2\linewidth}
    		\centering
    		\includegraphics[width=\textwidth,height=2cm]{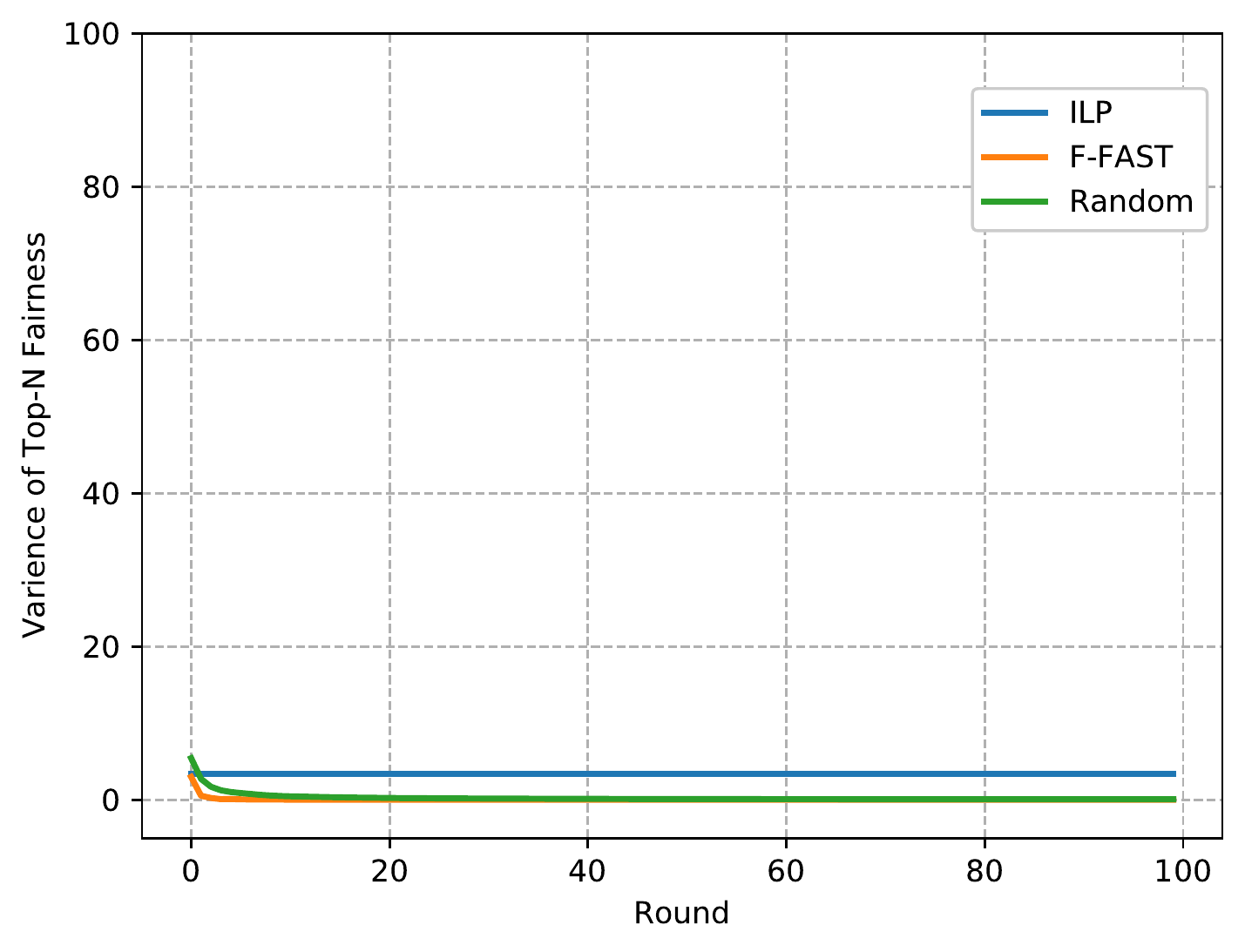}
    	\end{minipage}%
    }%
    \subfigure[$N = 5$]{
    	\begin{minipage}[t]{0.2\linewidth}
    		\centering
    		\includegraphics[width=\textwidth,height=2cm]{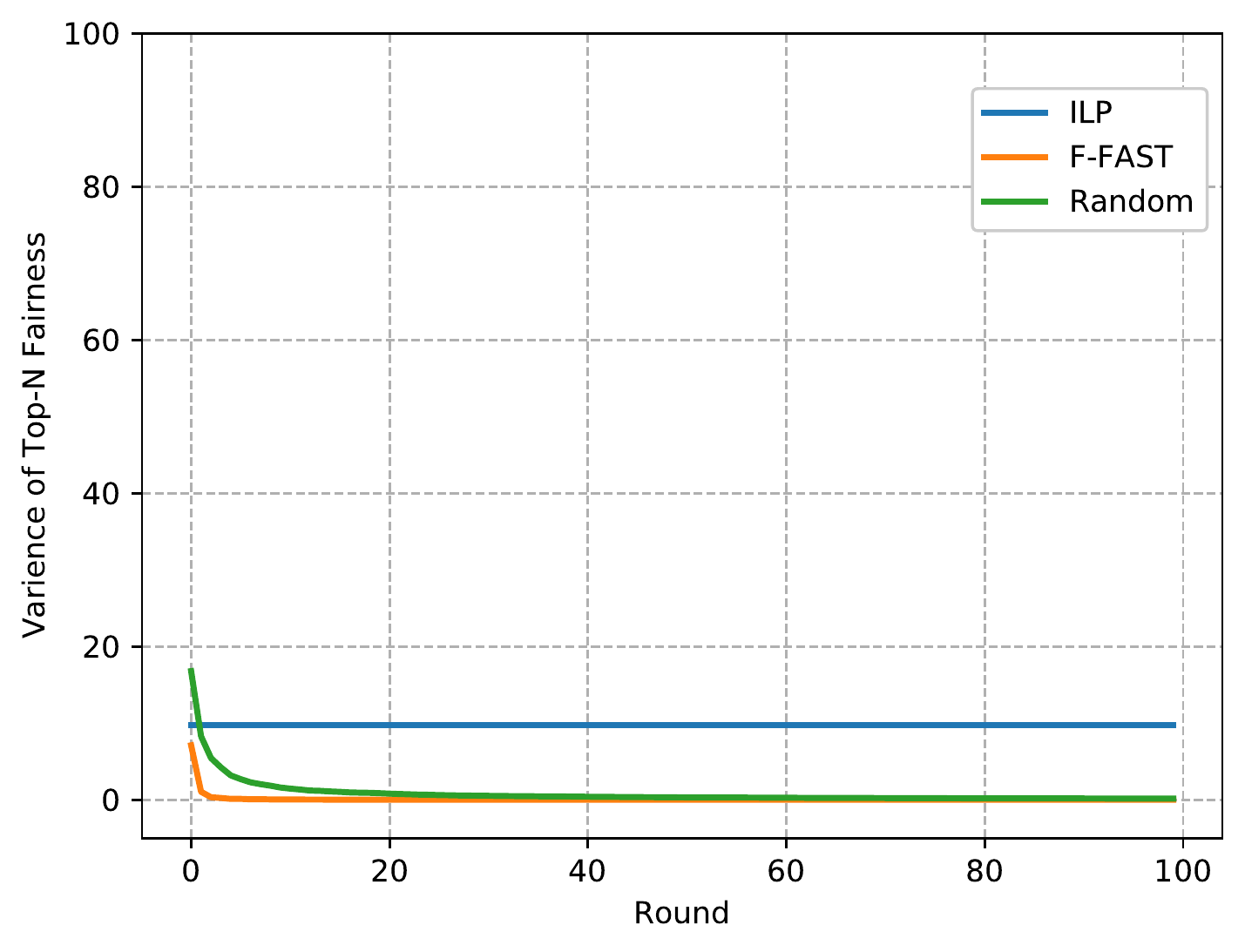}
    	\end{minipage}%
    }%
    \subfigure[$N = 10$]{
    	\begin{minipage}[t]{0.2\linewidth}
    		\centering
    		\includegraphics[width=\textwidth,height=2cm]{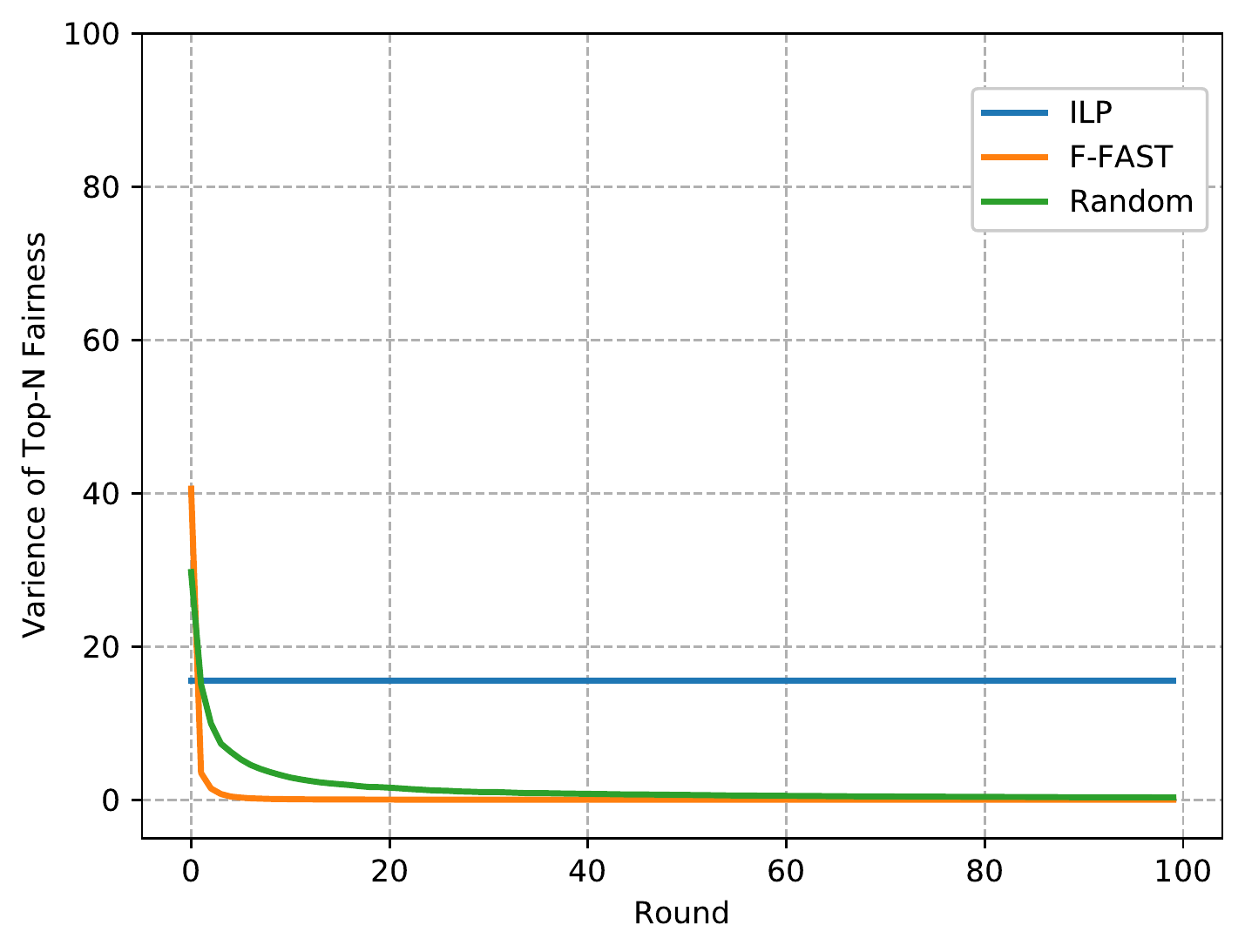}
    	\end{minipage}
    }%
    \subfigure[$N = 15$]{
    	\begin{minipage}[t]{0.2\linewidth}
    		\centering
    		\includegraphics[width=\textwidth,height=2cm]{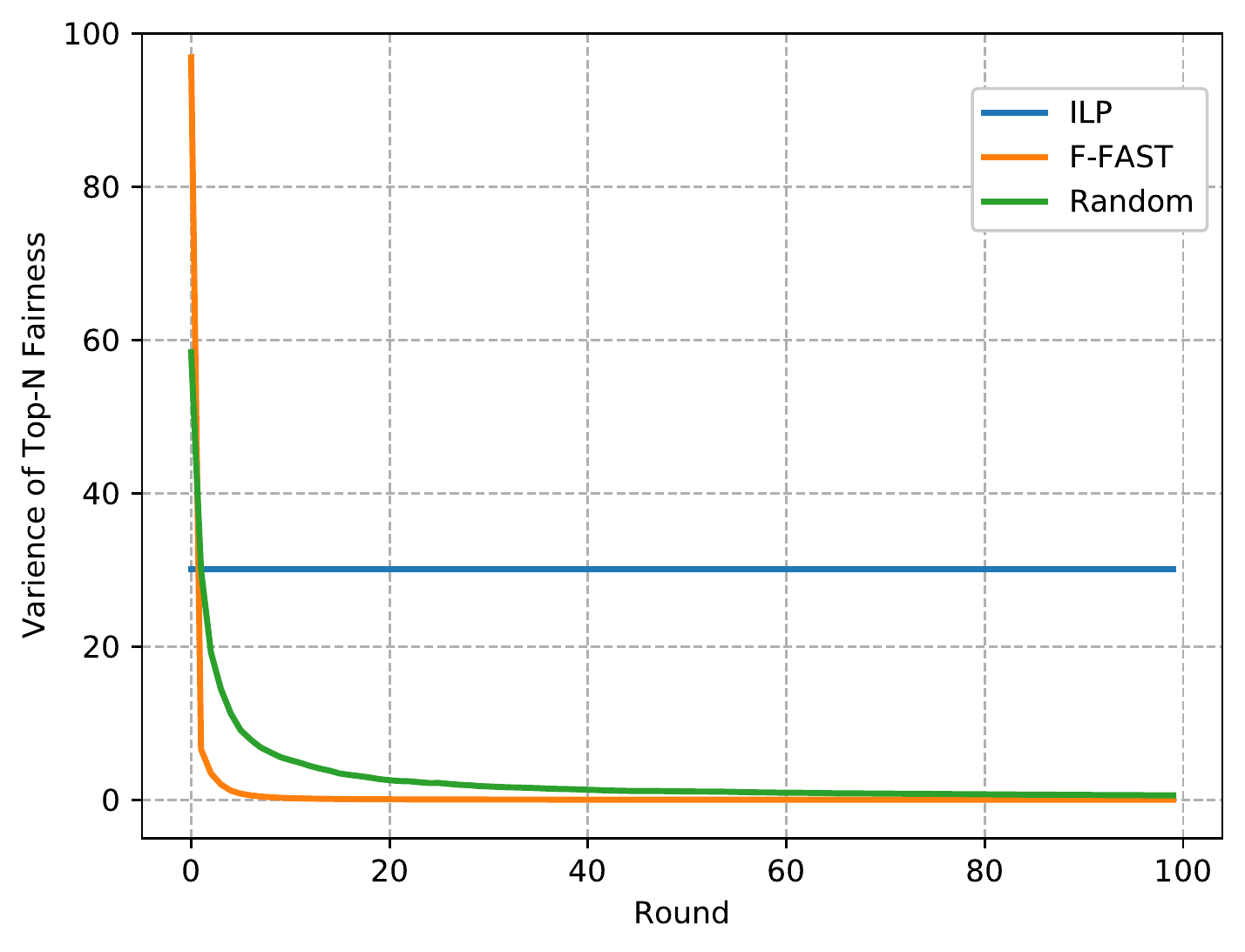}
    	\end{minipage}
    }%
    \subfigure[$N = 20$]{
    	\begin{minipage}[t]{0.2\linewidth}
    		\centering
    		\includegraphics[width=\textwidth,height=2cm]{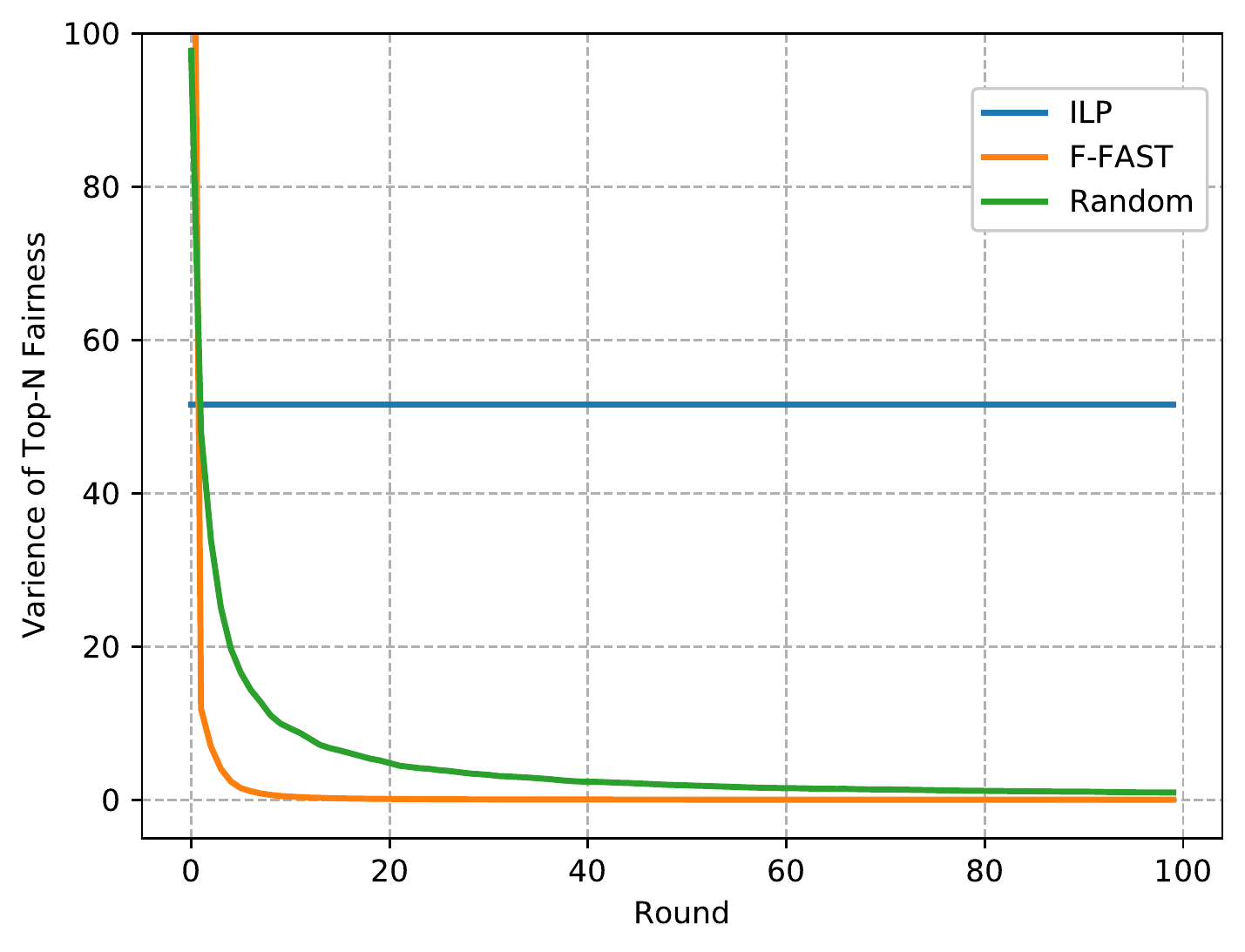}
    	\end{minipage}
    }%
	\centering
	\caption{Recommendation quality and variance of \textit{Top-N Fairness} under Different \textit{N}}
	\label{fig7}
\end{figure*}

Figure \ref{fig7} shows the performance of algorithms under different $N$. This experiment is performed on \textbf{Synthetic Dataset 2} under the premise of users being fixed, and a total of 100 rounds of recommendations are carried out for each experiment. 

As can be seen from the figures, as \textit{N} rises, the overall recommendation quality improves, but the rate of growth continues to decrease. It is worth noting that when \textit{N} is greater than 10, the quality of ILP and \textit{F-FAST} virtually does not increase and the quality of the random strategy even starts to decline. The reason for this is the impacts of the services in the lower positions in the original recommendation list is smaller. It can be noted that as \textit{N} rises, the variance of \textit{Top-N Fairness} rises instead. This shows that for \textit{F-FAST}, a longer top-$N$ recommendation list will be of benefit to recommendation quality but will impair fairness.

\subsubsection{Comparisons between Different Degrees of User Dynamics}
\begin{figure*}[!h]
	\centering
	\subfigure[20\%]{
		\begin{minipage}[t]{0.25\linewidth}
			\centering
			\includegraphics[width=\textwidth,height=2cm]{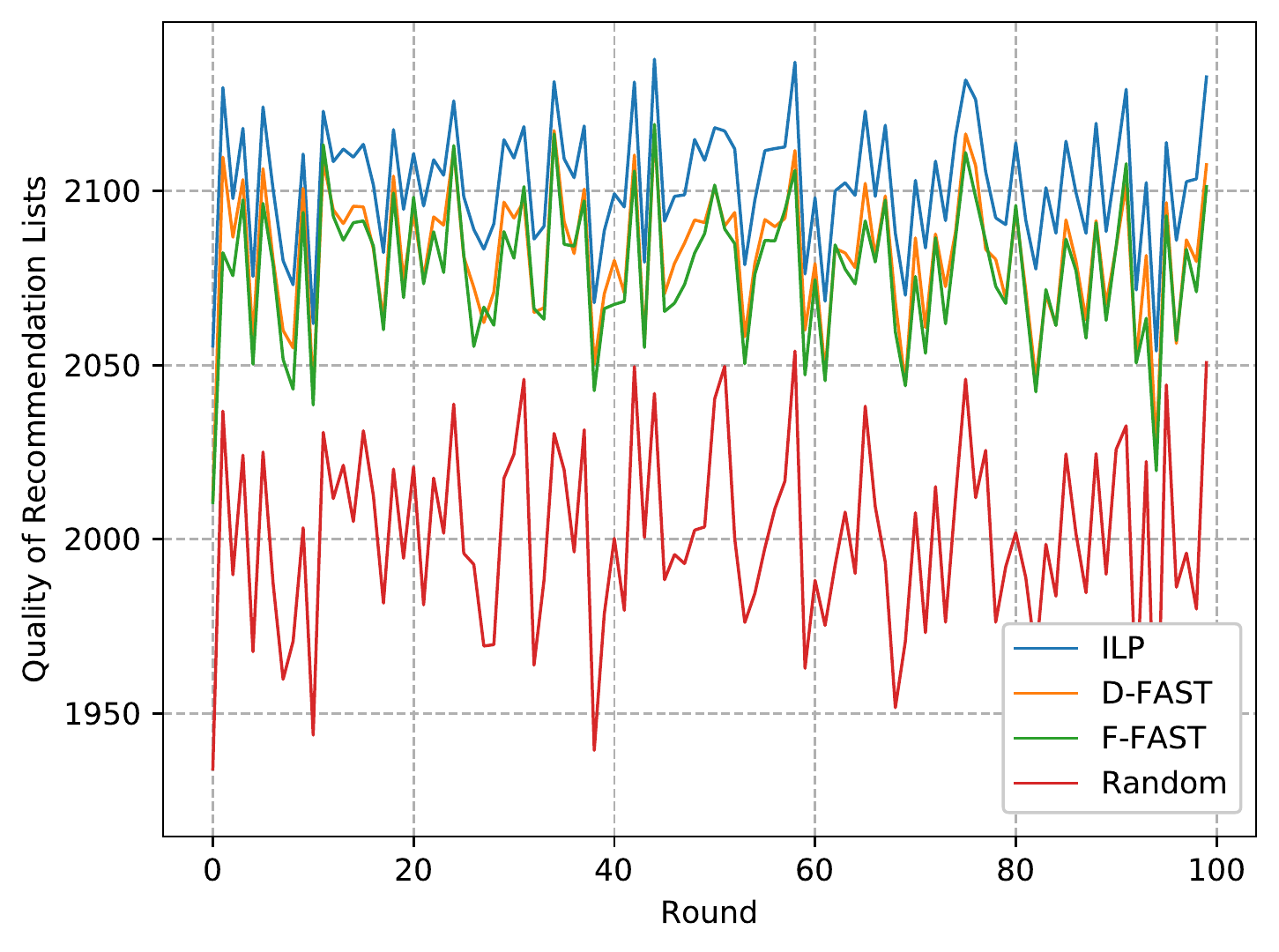}
		\end{minipage}%
	}%
	\subfigure[40\%]{
		\begin{minipage}[t]{0.25\linewidth}
			\centering
			\includegraphics[width=\textwidth,height=2cm]{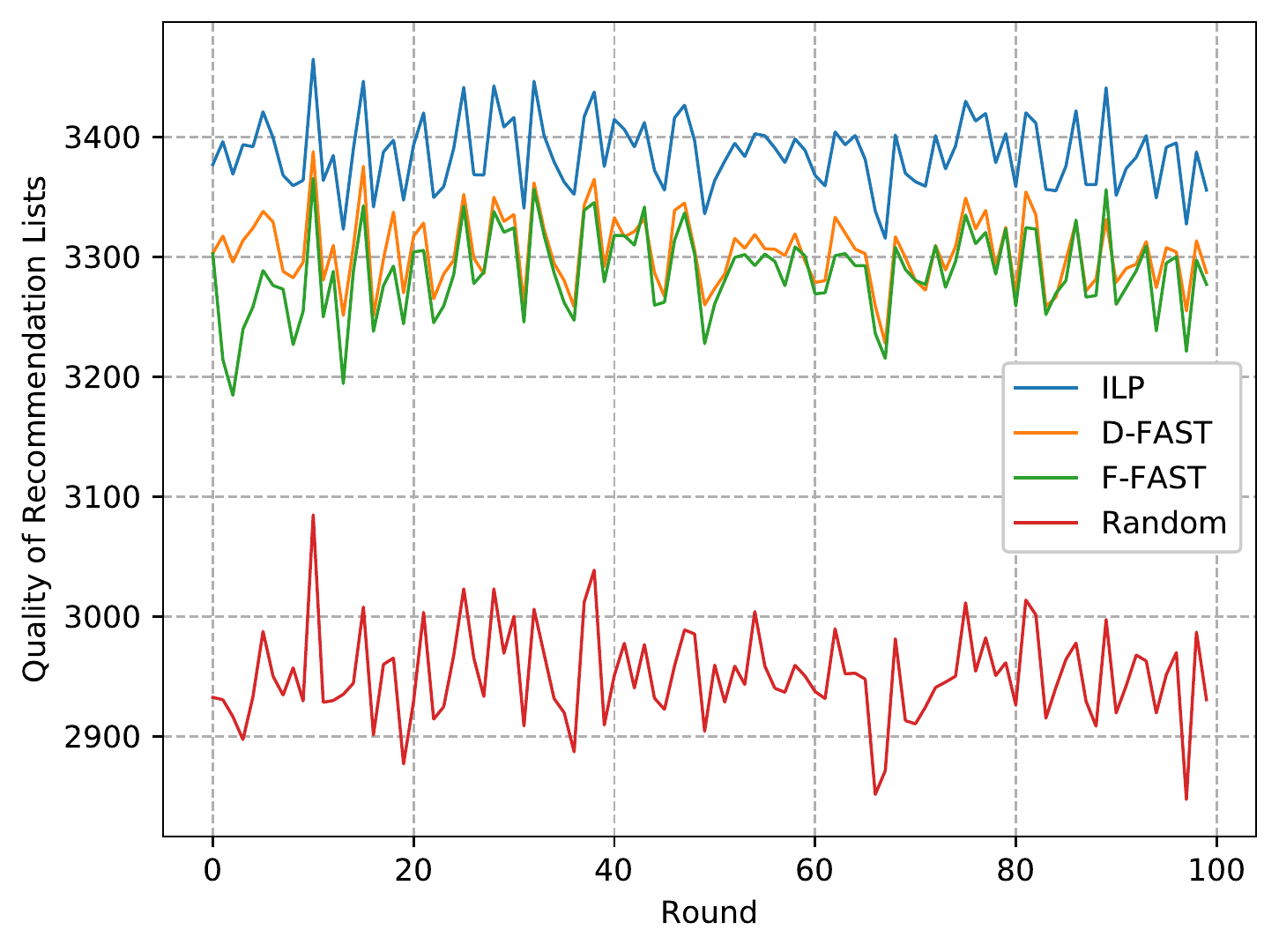}
		\end{minipage}%
	}%
	\subfigure[60\%]{
		\begin{minipage}[t]{0.25\linewidth}
			\centering
			\includegraphics[width=\textwidth,height=2cm]{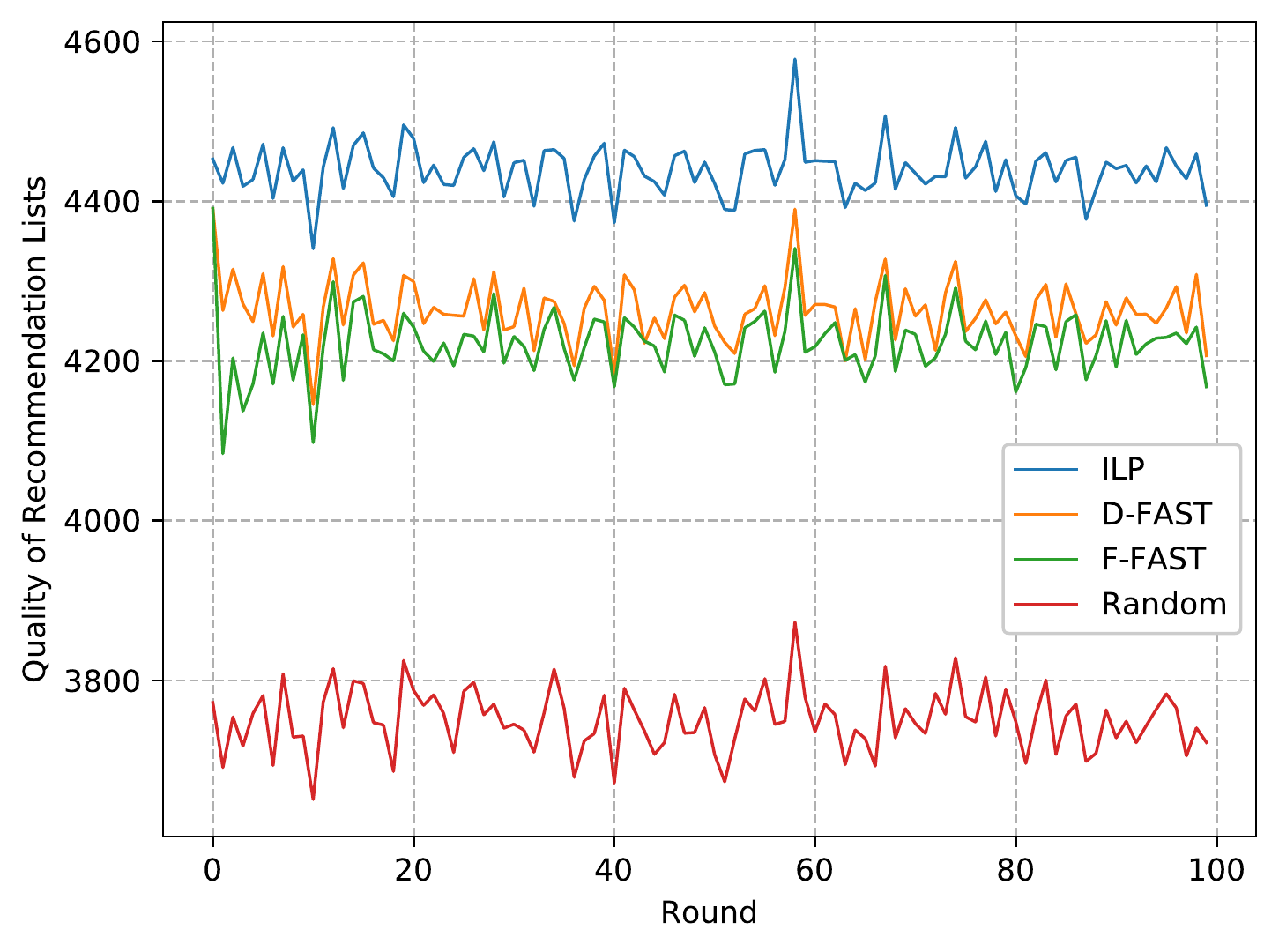}
		\end{minipage}
	}%
	\subfigure[80\%]{
		\begin{minipage}[t]{0.25\linewidth}
			\centering
			\includegraphics[width=\textwidth,height=2cm]{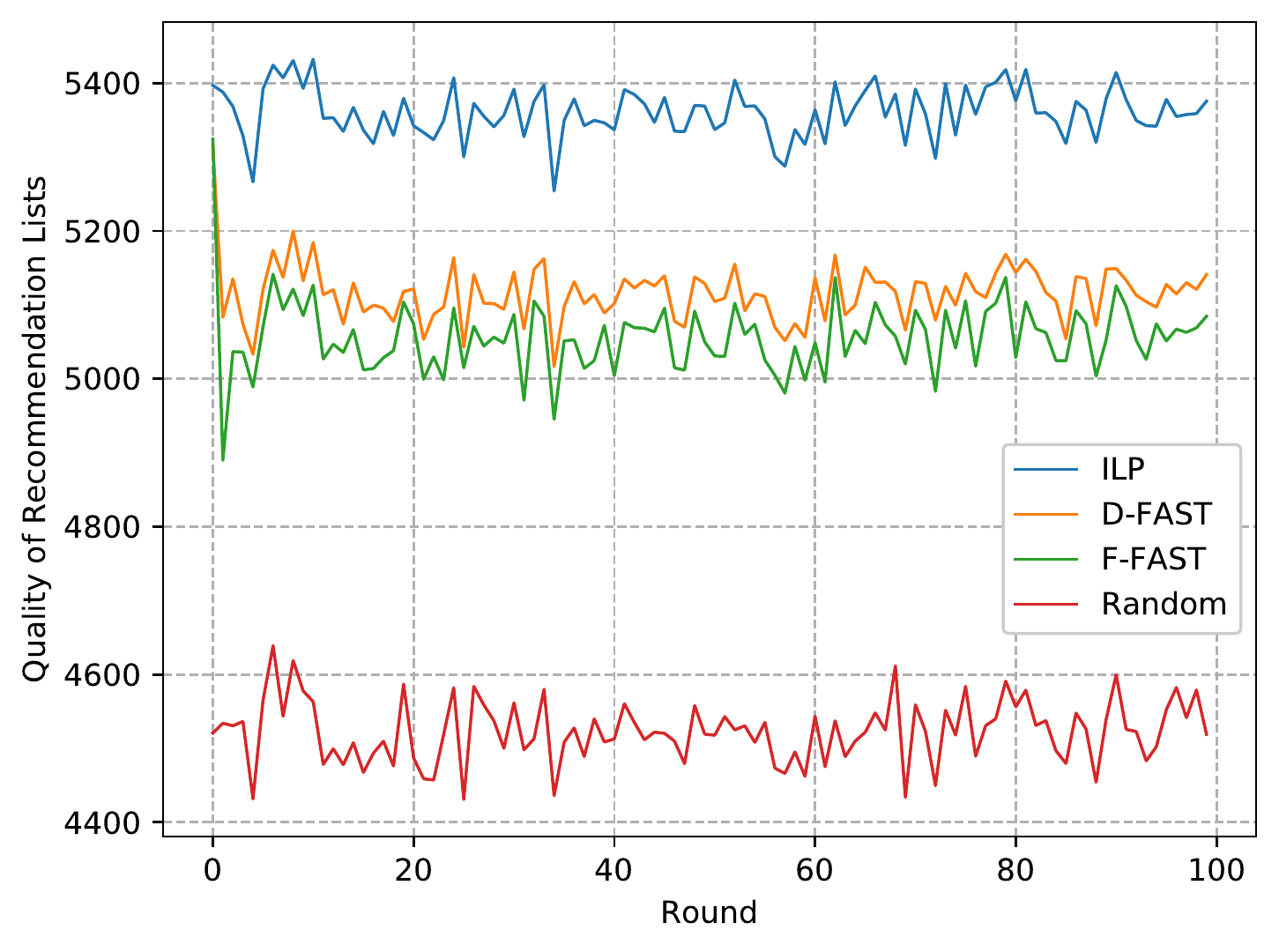}
		\end{minipage}
	}%
    \quad
    \subfigure[20\%]{
    \begin{minipage}[t]{0.25\linewidth}
    	\centering
    	\includegraphics[width=\textwidth,height=2cm]{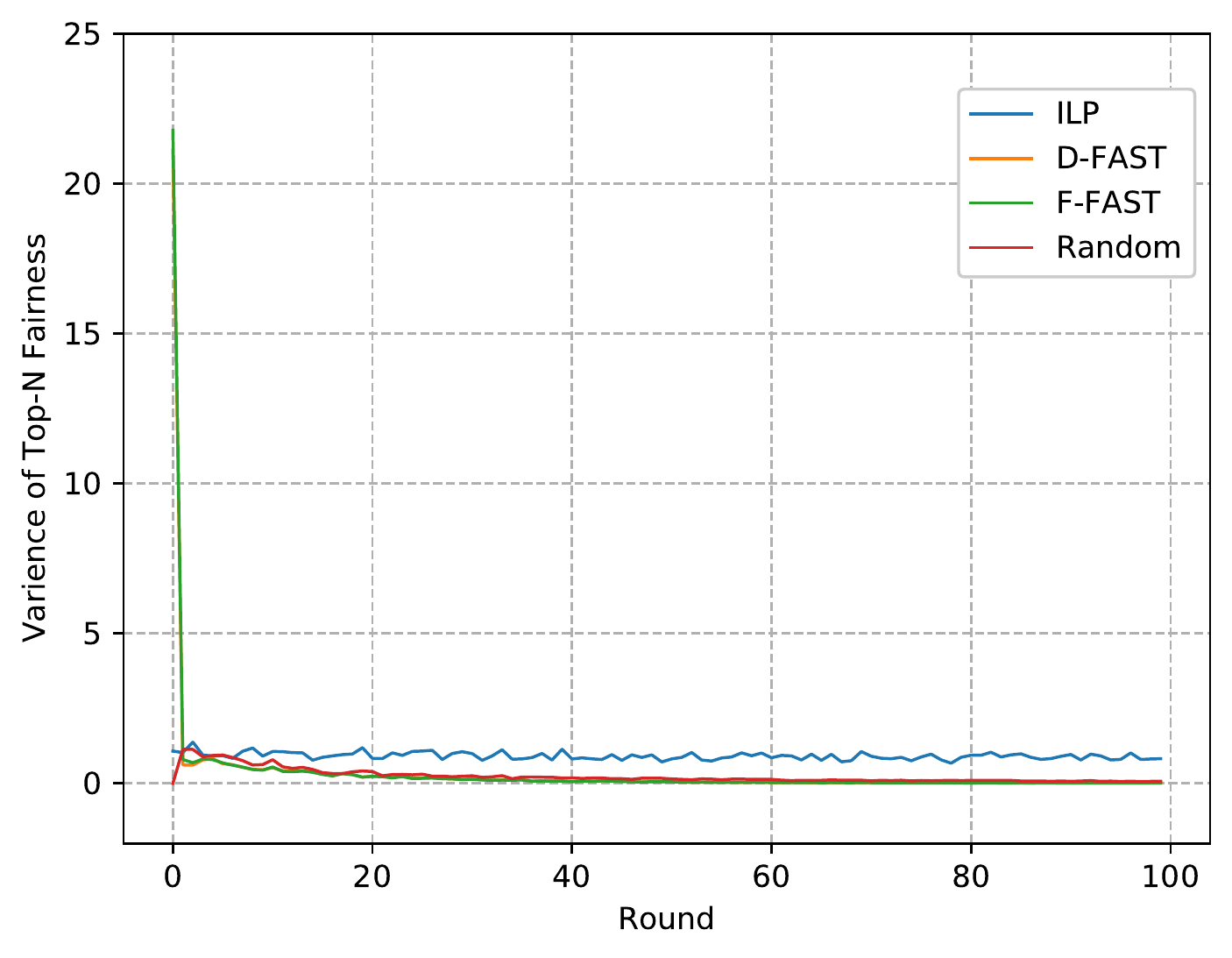}
    \end{minipage}%
    }%
    \subfigure[40\%]{
    	\begin{minipage}[t]{0.25\linewidth}
    		\centering
    		\includegraphics[width=\textwidth,height=2cm]{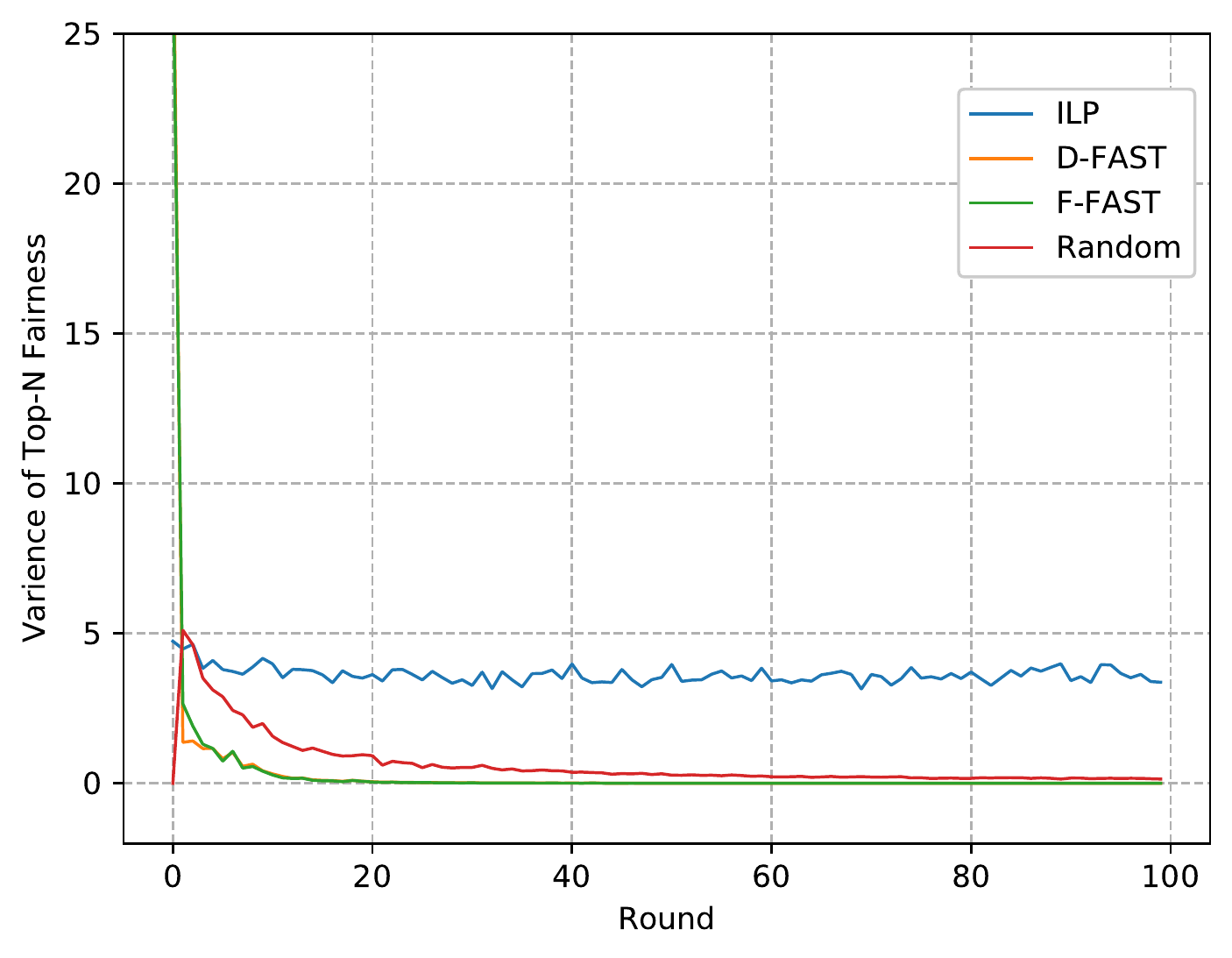}
    	\end{minipage}%
    }%
    \subfigure[60\%]{
    	\begin{minipage}[t]{0.25\linewidth}
    		\centering
    		\includegraphics[width=\textwidth,height=2cm]{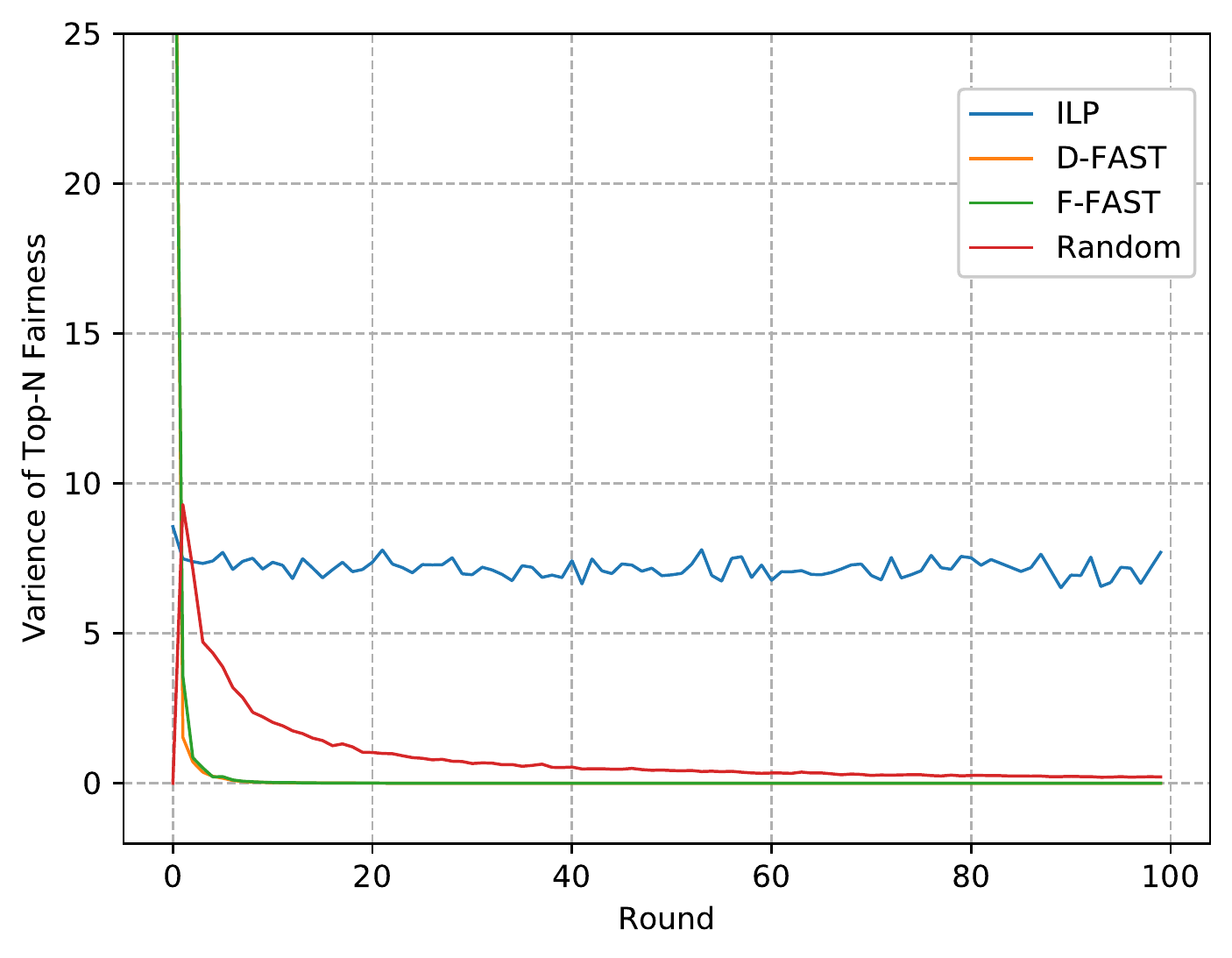}
    	\end{minipage}
    }%
    \subfigure[80\%]{
    	\begin{minipage}[t]{0.25\linewidth}
    		\centering
    		\includegraphics[width=\textwidth,height=2cm]{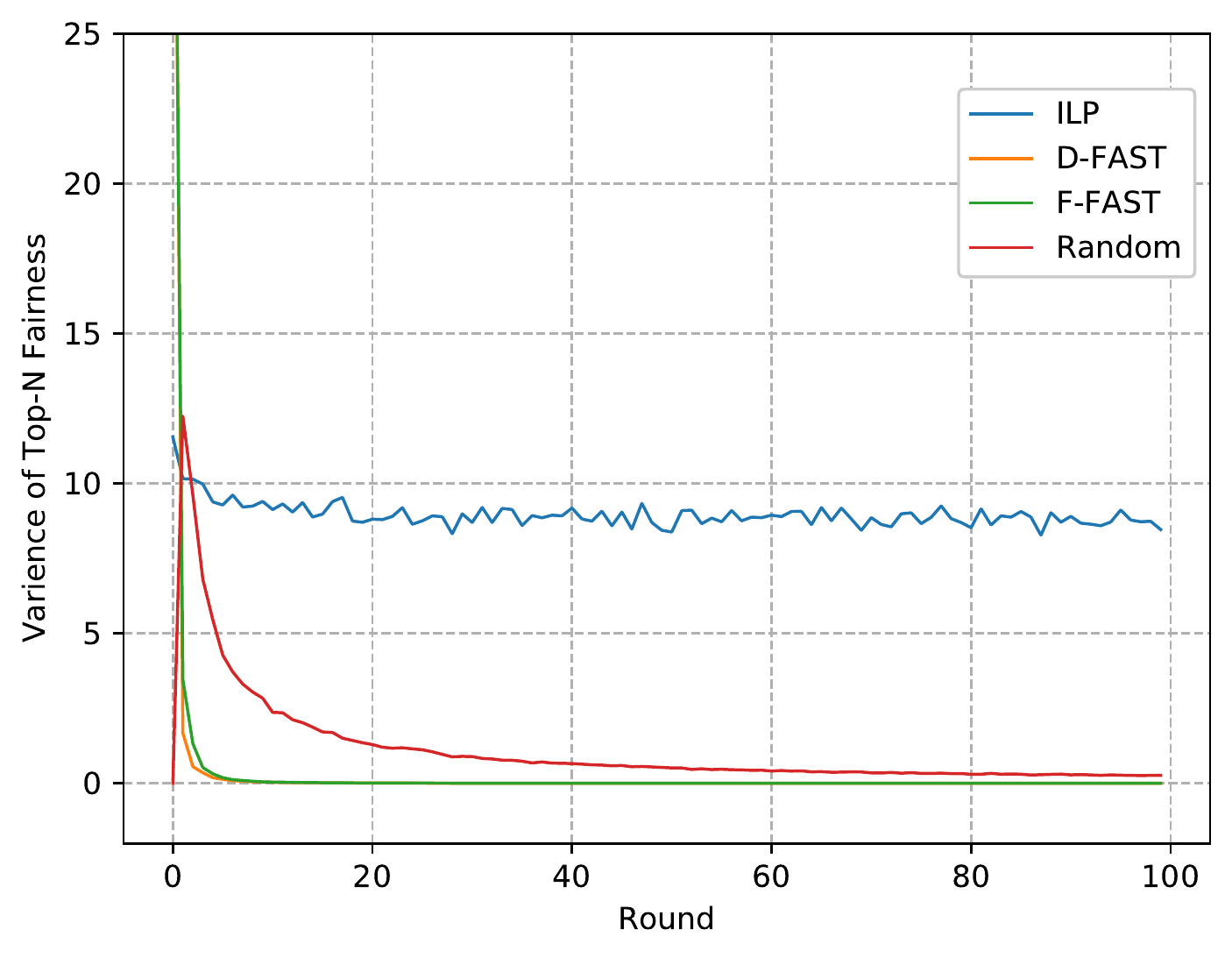}
    	\end{minipage}
    }%
	\centering
	\caption{Recommendation Quality and variance of \textit{Top-N Fairness} under Different Degrees of User Dynamics}
	\label{fig9}
\end{figure*}

We simulate the performance of the algorithms under different user dynamics by changing the proportion of recommended users in each round. Figures \ref{fig9} shows the results of the experiment on \textbf{Synthetic Dataset 2}.

It can be seen that both \textit{F-FAST} and \textit{D-FAST} perform very well on the dynamic user set with little loss of recommendation quality, and the variance of \textit{Top-N Fairness} is still close to zero. It is worth noting that \textit{D-FAST} can achieve lower fairness variance and higher recommendation quality than \textit{F-FAST}, which validates the measure of updating the \textit{Top-N Fairness} of each user according to their actual situations in \textit{D-FAST} which indeed improves the performance.

\subsubsection{Fairness of New Users}
\begin{figure}[!htbp]
	\centering
	\includegraphics[width=0.3\linewidth,height=3cm]{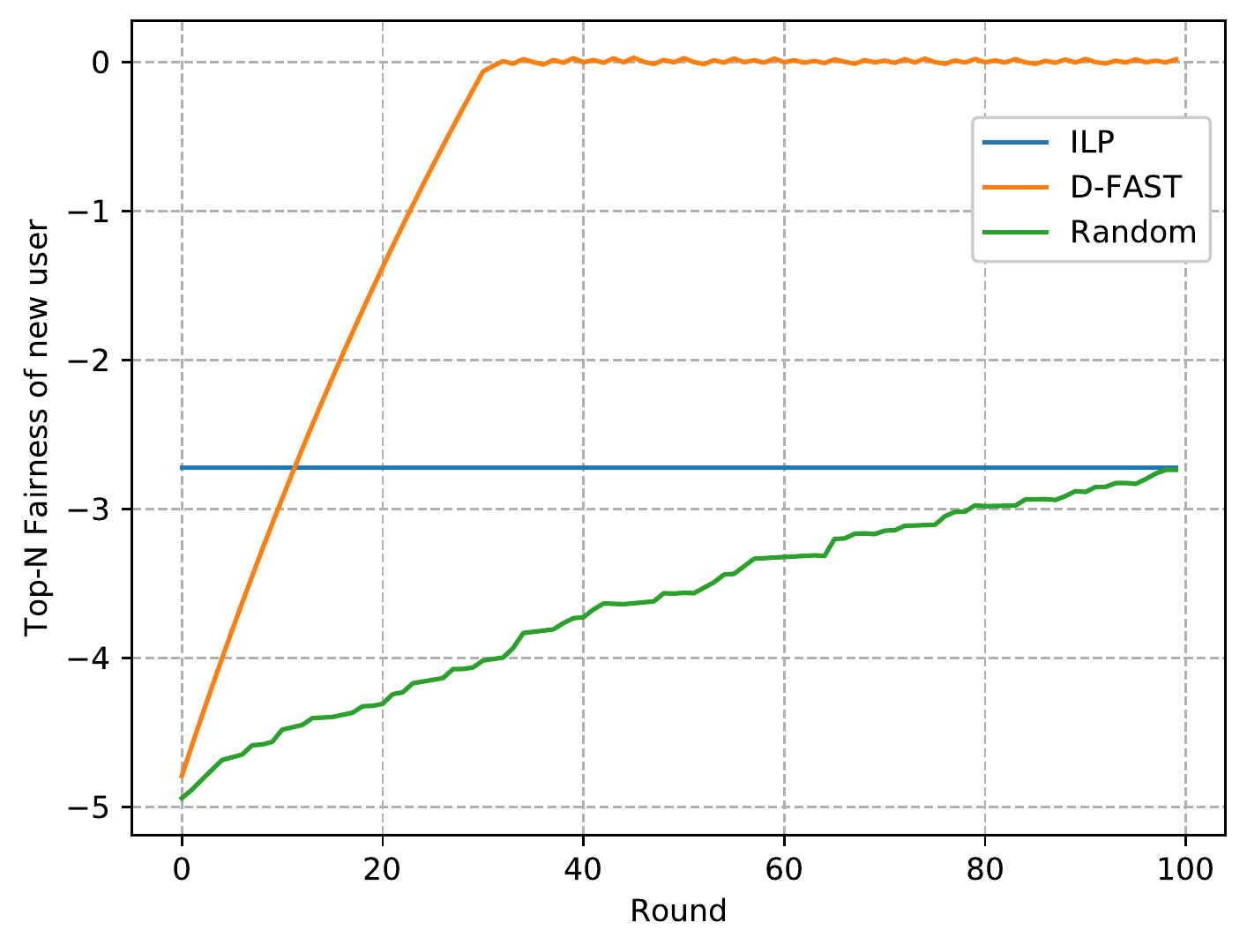}
	\caption{Trend of \textit{Top-N Fairness} for a New User} 
	\label{fig11}
\end{figure}

We compare how quickly the approaches can ensure a new user reaches a relatively fair state. We add a new user to a stable recommendation environment on \textbf{Synthetic Dataset 2} after 100 rounds of recommendations. We apply these four methods to the new recommendation environment after adding a new user, and compare the performance of the four methods on the \textit{Top-N Fairness} of the new user. The results are shown in Figure \ref{fig11}. Since the user set remains the same after adding new users, the process and results of \textit{F-FAST} and \textit{D-FAST} are the same, so we only show the results of \textit{D-FAST}.

As can be seen, \textit{D-FAST} ensures the new user reaches a relatively fair state (\textit{Top-N Fairness} being close to 0) in about 30 rounds of recommendation and this status is maintained. In contrast, although the random strategy can also ensure the \textit{Top-N Fairness} of a new user continues to approach 0, it is much slower. The ILP method cannot improve the fairness of new users.

\section{Conclusions}
This paper discusses the contradiction between the quality of recommendations and the fairness of users under the constraints of service capacity. We mainly consider fairness at the individual level, that is, to provide users with recommendations of equal quality, and propose a novel fairness measure \textit{Top-N Fairness} under the premise of capacity constraints. Based on this new metric, we design two heuristic algorithms for different user situations to resolve the contradiction. Through theoretical proofs and experiments, we verify that the proposed algorithms can ensure users reach a fair state while only sacrificing a small degree of recommendation quality. Going ahead, We want to extend individual fairness to group fairness and carry out relevant experimental and theoretical research.

\subsection*{Acknowledgement}
This work is partially supported by National Key Research and Development Plan(No. 2018YFB1003800).

%
%
\bibliographystyle{splncs04}
\bibliography{ref}

\begin{thebibliography}{10}
\providecommand{\url}[1]{\texttt{#1}}
\providecommand{\urlprefix}{URL }
\providecommand{\doi}[1]{https://doi.org/#1}

\bibitem{challenge2019yelp}
Yelp dataset challenges. \url{https://www.yelp.com/dataset/challenge/},
  accessed December 18, 2019

\bibitem{agarwal2018reductions}
Agarwal, A., Beygelzimer, A., Dud{\'\i}k, M., Langford, J., Wallach, H.: A
  reductions approach to fair classification. arXiv preprint arXiv:1803.02453
  (2018)

\bibitem{asudeh2019designing}
Asudeh, A., Jagadish, H., Stoyanovich, J., Das, G.: Designing fair ranking
  schemes. In: Proceedings of the 2019 International Conference on Management
  of Data. pp. 1259--1276. ACM (2019)

\bibitem{beutel2019fairness}
Beutel, A., Chen, J., Doshi, T., Qian, H., Wei, L., Wu, Y., Heldt, L., Zhao,
  Z., Hong, L., Chi, E.H., et~al.: Fairness in recommendation ranking through
  pairwise comparisons. arXiv preprint arXiv:1903.00780  (2019)

\bibitem{biega2018equity}
Biega, A.J., Gummadi, K.P., Weikum, G.: Equity of attention: Amortizing
  individual fairness in rankings. In: The 41st International ACM SIGIR
  Conference on Research \& Development in Information Retrieval. pp. 405--414.
  ACM (2018)

\bibitem{bose2019compositional}
Bose, A.J., Hamilton, W.: Compositional fairness constraints for graph
  embeddings. arXiv preprint arXiv:1905.10674  (2019)

\bibitem{burke2017multisided}
Burke, R.: Multisided fairness for recommendation. arXiv preprint
  arXiv:1707.00093  (2017)

\bibitem{calmon2017optimized}
Calmon, F., Wei, D., Vinzamuri, B., Ramamurthy, K.N., Varshney, K.R.: Optimized
  pre-processing for discrimination prevention. In: Advances in Neural
  Information Processing Systems. pp. 3992--4001 (2017)

\bibitem{christakopoulou2017recommendation}
Christakopoulou, K., Kawale, J., Banerjee, A.: Recommendation with capacity
  constraints. In: Proceedings of the 2017 ACM on Conference on Information and
  Knowledge Management. pp. 1439--1448 (2017)

\bibitem{ge2010beyond}
Ge, M., Delgado-Battenfeld, C., Jannach, D.: Beyond accuracy: evaluating
  recommender systems by coverage and serendipity. In: Proceedings of the
  fourth ACM conference on Recommender systems. pp. 257--260. ACM (2010)

\bibitem{geyik2019fairness}
Geyik, S.C., Ambler, S., Kenthapadi, K.: Fairness-aware ranking in search \&
  recommendation systems with application to linkedin talent search. arXiv
  preprint arXiv:1905.01989  (2019)

\bibitem{joachims2007evaluating}
Joachims, T., Granka, L., Pan, B., Hembrooke, H., Radlinski, F., Gay, G.:
  Evaluating the accuracy of implicit feedback from clicks and query
  reformulations in web search. ACM Transactions on Information Systems (TOIS)
  \textbf{25}(2), ~7 (2007)

\bibitem{joachims2007search}
Joachims, T., Radlinski, F.: Search engines that learn from implicit feedback.
  Computer  \textbf{40}(8),  34--40 (2007)

\bibitem{kamiran2012data}
Kamiran, F., Calders, T.: Data preprocessing techniques for classification
  without discrimination. Knowledge and Information Systems  \textbf{33}(1),
  1--33 (2012)

\bibitem{karako2018using}
Karako, C., Manggala, P.: Using image fairness representations in
  diversity-based re-ranking for recommendations. In: Adjunct Publication of
  the 26th Conference on User Modeling, Adaptation and Personalization. pp.
  23--28. ACM (2018)

\bibitem{liu2018personalizing}
Liu, W., Burke, R.: Personalizing fairness-aware re-ranking. arXiv preprint
  arXiv:1809.02921  (2018)

\bibitem{mehrotra2018towards}
Mehrotra, R., McInerney, J., Bouchard, H., Lalmas, M., Diaz, F.: Towards a fair
  marketplace: Counterfactual evaluation of the trade-off between relevance,
  fairness \& satisfaction in recommendation systems. In: Proceedings of the
  27th acm international conference on information and knowledge management.
  pp. 2243--2251 (2018)

\bibitem{patro2020incremental}
Patro, G.K., Chakraborty, A., Ganguly, N., Gummadi, K.P.: Incremental fairness
  in two-sided market platforms: On smoothly updating recommendations. AAAI,
  Feb  (2020)

\bibitem{pu2011user}
Pu, P., Chen, L., Hu, R.: A user-centric evaluation framework for recommender
  systems. In: Proceedings of the fifth ACM conference on Recommender systems.
  pp. 157--164. ACM (2011)

\bibitem{qian2015scram}
Qian, S., Cao, J., Mou{\"e}l, F.L., Sahel, I., Li, M.: Scram: A sharing
  considered route assignment mechanism for fair taxi route recommendations.
  In: Proceedings of the 21th ACM SIGKDD International Conference on Knowledge
  Discovery and Data Mining. pp. 955--964 (2015)

\bibitem{rastegarpanah2019fighting}
Rastegarpanah, B., Gummadi, K.P., Crovella, M.: Fighting fire with fire: Using
  antidote data to improve polarization and fairness of recommender systems.
  In: Proceedings of the Twelfth ACM International Conference on Web Search and
  Data Mining. pp. 231--239. ACM (2019)

\bibitem{stratigi2017fairness}
Stratigi, M., Kondylakis, H., Stefanidis, K.: Fairness in group recommendations
  in the health domain. In: 2017 IEEE 33rd International Conference on Data
  Engineering (ICDE). pp. 1481--1488. IEEE (2017)

\bibitem{yao2017beyond}
Yao, S., Huang, B.: Beyond parity: Fairness objectives for collaborative
  filtering. In: Advances in Neural Information Processing Systems. pp.
  2921--2930 (2017)

\bibitem{zafar2017fairness}
Zafar, M.B., Valera, I., Gomez~Rodriguez, M., Gummadi, K.P.: Fairness beyond
  disparate treatment \& disparate impact: Learning classification without
  disparate mistreatment. In: Proceedings of the 26th International Conference
  on World Wide Web. pp. 1171--1180. International World Wide Web Conferences
  Steering Committee (2017)

\bibitem{zehlike2017fa}
Zehlike, M., Bonchi, F., Castillo, C., Hajian, S., Megahed, M., Baeza-Yates,
  R.: Fa* ir: A fair top-k ranking algorithm. In: Proceedings of the 2017 ACM
  on Conference on Information and Knowledge Management. pp. 1569--1578. ACM
  (2017)

\bibitem{zhu2018fmsr}
Zhu, Q., Zhou, A., Sun, Q., Wang, S., Yang, F.: Fmsr: A fairness-aware mobile
  service recommendation method. In: 2018 IEEE International Conference on Web
  Services (ICWS). pp. 171--178. IEEE (2018)

\end{thebibliography}

\appendix
\section{Properties of \textit{F-FAST}}\label{Properties of F-FAST}
\textbf{THEOREM 1.} \emph{The sum of Top-N Fairness of all the users in each round is equal to zero}, $\sum_{u_i\in U}F_i^T = 0$.

\textbf{PROOF.} 
According to Equation (\ref{E4}), the sum of \textit{Top-N Fairness} can be formulated as:
\begin{equation}\label{E5}
\sum_{u_i\in U}F_i^T = \sum_{u_i\in U}\sum_{s_j\in l(N)_i}\frac{p_{i,j}^T-p^{T}_j}{p^{T}_j} = \sum_{s_j\in S}\sum_{u_i\in U_j}\frac{p_{i,j}^t-p^{T}_j}{p^{T}_j}
\end{equation}

Since all users will receive recommendations in every round, $\delta_i^t$ will all be equal to 1, and according to Equation (\ref{E2}) and Equation (\ref{E3}), $p^{T}_j$ and $p_{i,j}^t$ can be re-expressed as:
\begin{equation}\label{E13}
p^{T}_j = \frac{\sum_{u_i\in U_j}\sum^T_{t=0}In\_tn(s_j,l_i^t,N)}{\sum_{u_i\in U_j}T}, 
p_{i,j}^T = \frac{\sum^T_{t=0}In\_tn(s_j,l_i^t,N)}{T}
\end{equation}

Then the sum of \textit{Top-N Fairness} will be:
\begin{equation}\label{E7}
\begin{split}
\sum_{u_i\in U}F_i^T = &\sum_{s_j\in S}\sum_{u_i\in U_j}\frac{\frac{\sum^T_{t=0}In\_tn(s_j,l_i^t,N)}{T}}{\frac{\sum_{u_i\in U_j}\sum^T_{t=0}In\_tn(s_j,l_i^t,N)}{\sum_{u_i\in U_j}T}}
- \sum_{s_j\in S}\sum_{u_i\in U_j}1
\end{split}
\end{equation}

After reducing Equation (\ref{E7}), we can get:
\begin{equation}
\sum_{u_i\in U}F_i^T = \sum_{s_j\in S}\left(\frac{\sum_{u_i\in U_j}T}{T} - \sum_{u_i\in U_j}1\right) = 0
\end{equation}

\textbf{THEOREM 2.} \emph{Variance among Top-$N$ fairness of all users $D(F_i^T)$ converges to 0 with the recommended round $T$.}

\textbf{PROOF.}
According to \textbf{THEOREM 1}, we can get the mean of \textit{Top-N Fairness} of all users equals to zero, and the variance can be formulated as:
\begin{equation}\label{E10}
D(F_i^T) = \frac{\sum_{u_i\in U}\left(F_i^T \right)^2}{n}
\end{equation}

According to Equation (\ref{E4}), we know that:
\begin{equation}\label{E11}
\sum_{u_i\in U}\left(F_i^T \right)^2 = \sum_{u_i\in U}\left(\sum_{s_j\in l(N)_i}\frac{p_{i,j}^T-p^{T}_j}{p^{T}_j} \right)^2
\end{equation}

Since every user receives a recommendation in each round, $p^{T}_j$ of each service should be a constant. We discuss this issue in the following two cases.

For services without capacity conflicts: $c_j\geqslant len(U_j)$. Each service $s_j$ can be assigned to every user in its $U_j$ in each round. So $p^{T}_j$ and $p_{i,j}^T$ will always be 1. So the addends in summation formula of \textit{Top-N Fairness} are always equal to $0$ and can be ignored in this discussion.

For services with capacity conflicts: $c_j < len(U_j)$. Each service $s_j$ will always be assigned to $c_j$ users, so $p^{T}_j$ will be a constant less than 1, and we call it $Const_j$:
\begin{equation}
p^{T}_j ={c_j}/{len(U_i)} = Const_j < 1
\end{equation}

Then, Equation (\ref{E11}) will be:
\begin{equation}\label{E12}
\sum_{u_i\in U}\left(F_i^T \right)^2 = \sum_{u_i\in U}\left[\sum_{s_j\in l(N)_i}\left(\frac{p_{i,j}^T}{Const_j} -1\right)\right]^2
\end{equation}

The only variable in Equation (\ref{E12}) is $p_{i,j}^t$, and according to Equation (\ref{E13}), we can get:
\begin{equation}
\begin{split}
p_{i,j}^{T+1} &= \frac{\sum^T_{t=0}In\_tn(s_j,l_i^t,N)}{T+1} + \frac{In\_tn(s_j,l_i^{T+1},N)}{T+1}
\end{split}
\end{equation}

According to \textbf{THEOREM 1}, we can divide users into two groups, users with low \emph{Top-N Fairness}($F_i^T < 0$) and users with high \emph{Top-N Fairness}($F_i^T \geqslant 0$).

For users with low \emph{Top-N Fairness}, addends with $p_{i,j}^T < Const_j$ occupy the main influence factor in the summation formula of \emph{Top-N Fairness} in this situation. As designed in our strategy, users with low \emph{Top-N Fairness} will always be allotted first, that:
\begin{equation}
\begin{split}
1>Const_j &> P_{i,j}^{T+1} = \frac{\sum^T_{t=0}In\_tn(s_j,l_i^t,N) + 1}{T+1} > \frac{\sum^T_{t=0}In\_tn(s_j,l_i^t,N)}{T} = P_{i,j}^T
\end{split}
\end{equation}

According to Equation (\ref{E12}), we know that
\begin{equation}
\left|F_i^{T+1}\right| < \left|F_i^T\right|, (F_i^{T+1})^2 < (F_i^T)^2
\end{equation}

For users with high \emph{Top-N Fairness}, addends with $P_{i,j}^T \geqslant Const_j$ occupy the main influence factor in the summation formula of \emph{Top-N Fairness} in this situation. Also, according to our recommendation strategy, these users will always be assigned last and will most likely not be assigned under the condition that service capacity is limited, so that:
\begin{equation}
\begin{split}
Const_j \leqslant p_{i,j}^{T+1} &= \frac{\sum^T_{t=0}In\_tn(s_j,l_i^t,N)}{T+1} < \frac{\sum^T_{t=0}In\_tn(s_j,l_i^t,N)}{T} = P_{i,j}^T
\end{split}
\end{equation}

According to Equation (\ref{E12}), we can also get:
\begin{equation}
\left|F_i^{T+1}\right| < \left|F_i^T\right|, (F_i^{T+1})^2 < (F_i^T)^2
\end{equation}

In both cases, $(F_i^T)^2$ becomes smaller as the round of recommendation increases. When a user's $F_i^T$ is not equal to the average $F_i^T$ of users, \textit{F-FAST} will continue to work until $F_i^T$ of all users is equal, and we can get that $D(F_i^T)$ will converge to 0, thus \textbf{THEOREM 2} is true.

\end{document}